\documentclass[aps,prev,preprintnumbers,floatfix,nofootinbib]{article}
\usepackage{macro_article}
\usepackage{graphics}
\usepackage{epstopdf}
\usepackage{hyperref}
\usepackage{mathrsfs}
\usepackage{amssymb}
\usepackage{verbatim}

\newcommand{\slpartial}{\raise.15ex\hbox{$/$}\kern-.53em\hbox{$\partial$}}
\newcommand{\slcalP}{\raise.15ex\hbox{$/$}\kern-.63em\hbox{$\cal P$}}
\newcommand{\slp}{\raise.15ex\hbox{$/$}\kern-.63em\hbox{$p$}}
\newcommand{\slk}{\raise.15ex\hbox{$/$}\kern-.63em\hbox{$k$}} 

\definecolor{greye}{rgb}{0.5,0.5,0.5}

\begin{document}

\vspace*{-.6in} \thispagestyle{empty}
\begin{flushright}
LPT--Orsay-13-49, 
CPHT-RR054.0613,
IFT-UAM/CSIC-13-76

\end{flushright}
\baselineskip = 20pt

\vspace{.5in} {\LARGE
\begin{center}
{\bf  Extra U(1), effective operators, anomalies and dark matter}
\end{center}}

%\vspace{1cm}
%\vspace{.5in}

\begin{center}
{\bf  Emilian Dudas$^{~a}$, Lucien Heurtier$^{~a}$, Yann Mambrini$^{~b}$ and   
Bryan Zaldivar$^{~c}$   }  \\

%\vspace{.5in}

$a$: \emph{CPhT, Ecole Polytechnique, 91128 Palaiseau Cedex, France
 }
\\ 
$b$: \emph{Laboratoire de Physique Th\'eorique, 
Universit\'e Paris-Sud, F-91405 Orsay, France  }
\\
$c$: \emph{Instituto de Fisica Teorica, IFT-UAM/CSIC,  28049 Madrid, Spain }

\end{center}

 \begin{abstract}
\noindent 
A general analysis is performed on the dimension-six operators mixing an almost
hidden $Z'$ to the Standard Model (SM), when the $Z'$ communicates with the SM
via heavy mediators. These are fermions charged under both $Z'$ and the SM, while all SM fermions are neutral under $Z'$. We classify the operators as a function of the gauge anomalies behaviour of mediators and explicitly compute the dimension-six operators coupling $Z'$ to gluons, generated at one-loop by chiral but anomaly-free, sets of fermion mediators.
We prove that only one operator contribute to the couplings between $Z'$ charged
matter and on-shell gluons. We then make a complete phenomenological analysis of the scenario 
where the lightest fermion charged under $Z'$ is the dark matter candidate.
Combining results from WMAP/PLANCK data, mono-jet searches at LHC, and direct/indirect dark matter detections restrict considerably the allowed parameter space.
\end{abstract}

\newpage

\setcounter{page}{1}

\tableofcontents

%\twocolumn[]

\newpage

%%%%%%%%%%%%%%%
%%%%%%%%%%%%%%%

\section{Introduction and Conclusions}

New abelian gauge symmetries are arguably the simplest extensions of the Standard Model (SM) (for a recent review, see \cite{zprime}) . 
If SM fermions are charged under a new abelian $U(1)_X$, its couplings are strongly constrained by direct searches and especially by FCNC processes. The simplest and widely studied
possibility in the literature is when SM fermions have flavor-independent charges. Most popular examples in this class are $B-L$ or linear combinations $\alpha (B-L) + \beta Y$. 
They are actually the only family-independent, anomaly-free gauged symmetries commuting with the SM gauge group in case where there are no new fermions beyond the 
ones of the SM. Family-dependent anomaly-free models with no extra fermions
were also extensively studied\footnote{For recent updates on phenomenological and experimental constraints on such models, see e.g. \cite{fabio}.}. In all such cases,
the $Z'$ should be heavy enough to escape detection, at least in the multi-TeV range.   
There is also a large literature on light $U(1)$'s of string or field theory origin with anomaly cancellation a la Green-Schwarz \cite{dumitru,abdk,coriano,wells}, with
low-energy anomalies canceled by axionic couplings and generalized Chern-Simons terms,
or in other models with Stueckelberg realization of $Z'$ \cite{nath}.  

A radically different option is to have no SM fermions charged under $Z'$. This is a relatively natural framework in string theory with D-branes. But it is also natural
from a field theory viewpoint, with additional heavy fermions $\Psi_{L,R}$, called ``mediators" in what follows, which mediate effective interactions, described 
by the dimension-four kinetic mixing and higher-dimensional operators between the $Z'$ and the SM sector \cite{dmpr}. If one wants mediators parametrically
heavier than the electroweak scale (say in the TeV range), we need, in addition to possible SM Higgs contributions, an additional source to their mass. 
A purely Dirac mass is of course a simple viable option. However as argued in \cite{dmpr}, because of the Furry theorem, the only low-dimensional  induced effective 
operator is the kinetic mixing, whereas the next higher-dimensional ones are of dimension eight.  Throughout our paper, we consider the kinetic mixing to be
small enough. If we are interested in $Z'$ couplings to gluons, this can be achieved  for example by having colored mediators with no hypercharge. 
In this case, the main couplings between the ``hidden" $Z'$ and the SM are generated by higher-dimensional effective operators (hdo's), the lowest relevant 
ones being of dimension six.  However, we will show that in the parameter space allowed by the PLANCK/WMAP data, the  phenomenological consequences
induced by the presence of a kinetic mixing allowed by various constraints are negligible. 
The simplest and natural option to obtain dimension-six effective operators is to generate the mediator masses  by the vev of the scalar field $\phi$ breaking 
spontaneously the $Z'$ gauge symmetry. The corresponding induced mediator masses, called generically $M$ in what follows, determine the mass scale
 of the hdo's and also the UV cutoff of the effective theory. There could also be contributions to their mass from the SM Higgs field $m \sim \lambda \langle H \rangle = \lambda v$, 
which are considered to be smaller, such  that we can expand  in powers of $v/M$ and obtain operators invariant under the SM gauge group. Such a framework 
 was already investigated in
\cite{dmpr,tait} from the viewpoint of the effective couplings of $Z'$ to electroweak
gauge bosons. The potential implications to dark matter, considered to be the lightest 
fermion in the dark sector was also investigated, with the outcome that a monochromatic gamma ray line from the dark matter annihilation 
is potentially observable. The potential existence of a signal in the FERMI data was largely discussed in the recent literature \cite{weniger} and 
will not be discussed further here. 

In this paper we extend the previous works by allowing the mediators to be colored
and therefore the $Z'$ to couple to gluons. We restrict ourselves throughout the paper to CP even couplings for simplicity. These couplings are more restricted by symmetries than the ones to the electroweak gauge bosons and their presence change
significantly the phenomenology of such models. Whereas at dimension-six order four
such operators are possible, only two of them are induced by heavy fermion mediators loops. Moreover, only one operator contributes to amplitudes in which at least one of the gluons is on-shell, as will be the case throughout our paper. We analyze in detail the corresponding phenomenology from the viewpoint of
the dark matter relic abundance, direct and indirect dark matter detection and LHC constraints.  Allowing couplings to gluons and at the same time to electroweak gauge bosons does not change significantly the phenomenology of the $Z'$ compared to the case where only couplings
to gluons are allowed. One interesting conceptual difference is that, whereas the
$Z'$ couplings to gluons and photons {\it vanish} for an on-shell $Z'$ due to the Landau-Yang theorem \cite{landauyang}, the couplings to
 the electroweak gauge bosons $ZZ$,$Z \gamma$ do not vanish; they lead on the contrary to an {\it enhancement} close to the $Z'$ pole. Another interesting result is
  that, unlike the case of kinetic mixing, the dark matter annihilation into gluons induced by virtual $Z'$ exchange can give correct relic density  for heavy dark matter 
  and $Z'$ masses, well above the electroweak scale. Since our interest here is to have complementary constraints from dark matter searches and LHC, we nonetheless 
  confine our analysis to masses below than or of the order TeV in what follows. 
 
The paper is organized as follows. 
Section 2 introduces the basic formalism we will use, which is Stueckelberg
realization of $Z'$ symmetry. It contains the list of the lowest dimensional effective operators generated by integrating-out heavy fermionic mediators, their classification 
depending on the nature of  messenger masses and charges and the explicit loop computation of the $Z'$ couplings to gluons. Section 3 deals with the consequences of the model for dark matter generation in the Early Universe, focusing on the annihilation to a gluon pair. Section 4 contains the 
various phenomenological constraints coming from the unique $Z'$ coupling to gluons generated at one-loop
by heavy colored mediators.  Section 5 contains the re-analysis of the various
constraints when $Z'$ couplings to electroweak gauge bosons are also added. 
Appendices contain more details about the gauge independence of the $Z'$ mediated hidden-sector-SM couplings, the effective operator couplings
 $Z'$ to gluons induced by heavy mediator loops and the complete cross-sections of the s- and t-channel annihilation of the dark matter. 
 
%%%%%%%%%%%%%%%%%%%%%%%%%%%%%%%%%%%%%%%%%%%%%%%%%%%%%%%%%%%%%%%%%%%%%%%%%%%%%
\section{$Z'$, heavy fermion mediators and effective operators}

 The effective lagrangian generated by loops of heavy mediators is generically invariant under SM and has a non-linear (Stueckelberg) realization for $Z'$, for the following reason. If the mediator masses are invariant
 under both the SM and the $Z'$ gauge symmetry, the induced operators would be gauge invariant in the usual sense. If the mediator masses are however generated by the
 breaking of $U(1)_X$, in the broken phase below the mass of the heavy Higgs $\phi$ breaking $U(1)_X$, the symmetry is still present but realized a la Stueckelberg.  Indeed,   in the limit where  $\phi$ is much heavier than the $Z'$, in the effective theory we keep only the axionic
component of the original $Z'$ Higgs field $\Phi = \frac{V + \phi}{\sqrt{2}} \exp(i a_X/V) \to \frac{V }{\sqrt{2}} \exp(i a_X/V) $.  We  define the 
dimensionless axion $\theta_X = \frac{a_X}{V}$ in what follows. The axion transforms
non-linearly under $U(1)$ transformations
\begin{equation}
\delta Z'_{\mu} \ = \ \partial_{\mu} \alpha \quad , \quad 
\delta \theta_X \ = \ \frac{g_X}{2} \ \alpha \ . \label{inv1}
\end{equation}

The exact lagrangian, describing all the microscopic physics, including the mediator fields $\Psi_{L,R}$, is then of the form 
\begin{eqnarray}
 {\cal L} & =&  {\cal L}_{SM} + \  {\bar \Psi}_L^{i} \left( i \gamma^{\mu}
\partial_{\mu} + \half{g_X} X^{i}_L \gamma^{\mu} Z_{\mu}' \right)
\Psi^{i}_L + {\bar \Psi}^{i}_R \left( i \gamma^{\mu}
\partial_{\mu} + \half{g_X} X^{i}_R \gamma^{\mu} Z_{\mu}' \right)
\Psi^{i}_R \nonumber \\
&-& \left( {\bar \Psi}^{i}_L M_{ij} e^{i a_X (X_L^i-X_R^j) \over V} \Psi^{i}_R
+ {\rm h.c.} \right) \  + \ {1 \over 2} (\partial_{\mu} a_X - M_{Z'} Z'_{\mu} )^{2} -
\frac{1}{4} F^X_{\mu \nu} F^{X \, \mu \nu}
\end{eqnarray}
where ${\cal L}_{SM}$ is the Standard Model Lagrangian and where $M_{Z'} = g_X V/2$. This lagrangian is indeed invariant under (\ref{inv1}), with non-linear shifts of the axion $a_X$ crucial
for restauring gauge invariance. 
If the original high-energy lagrangian is anomaly-free and the SM fermions are neutral
under $Z'$, then the mediators have to form an anomaly-free set. We are considering this class of models in most of this paper.  In this case, the induced effective operators are gauge invariant a la Stueckelberg. Throughout the paper we restrict ourselves to CP even operators for simplicity. In the case where the mediators are not an anomaly-free set, then either low-energy fermions have to be charged under $Z'$, or there are axionic couplings and GCS terms in order to cancel anomalies\footnote{A general field-theoretical analysis with computation of these couplings and analysis of anomalies cancellation can be found in \cite{abdk}.}.
For notational convenience we define:
\begin{eqnarray}
 {D}_{\mu} \theta_X \equiv \ \partial_{\mu} \theta_X  - \frac{g_X}{2} Z'_{\mu} \ , &&~~\widetilde{F}_{\mu \nu} \equiv \frac{1}{2} \epsilon_{\mu \nu \rho \sigma} F^{\rho \sigma}\,,\nonumber \\
%{\cal D}_{\mu} \theta_i &\equiv& \left(\partial_{\mu} \theta_i  - g' Z'_{\mu} \right) \nonumber \\
 {\cal T}r(F G) \equiv {\rm Tr} [ F_{\mu \nu} G^{\mu \nu} ]
\,\,  ,&& ~~{\cal T}r (E F G) \equiv {\rm Tr} [ E_{\mu}^{~\lambda} F_{\lambda \nu} G^{\nu \mu}] \ ,
\label{inv5}
\end{eqnarray}
where ${\rm Tr}$ takes into account a possible trace over non-abelian indices.
In summary, there are three distinct possibilities: 

\noindent
{\bf i)} The mediators are completely non-chiral, i.e. vector-like both respect to the
SM and $U(1)_X$. In this case, there are no dimension-six induced operators, since the
only one that can be potentially written, $\mathcal{T}r(F^X F_{SM} \tilde{F}_{SM})$ vanishes exactly as shown in the Appendix. \\
{\bf ii)} The mediators form an anomaly-free set, but are chiral with respect to $U(1)_X$ and vector-like with respect to the SM. The induced dimension-six operators in this
case are
\begin{eqnarray}
 \mathcal{L}^{(6)}_{\text{CP even}}&=&\frac{1}{M^2}\left\{ d_g\partial^{\mu} D_{\mu} \theta_X {\cal T}r(G\tilde{G}) + 
d'_g \partial^{\mu} D^{\nu} \theta_X {\rm Tr}(G_{\mu \rho}\tilde{G}^{\rho}_{\nu})  \nonumber
\right. \\
&+& \left. e_g  D^{\mu} \theta_X {\rm Tr}(G_{\nu \rho} {\cal D}_{\mu} \tilde{G}^{\rho \nu})
+ e'_g D_{\mu} \theta_X {\rm Tr}(G_{\alpha \nu} 
{\cal D}^{\nu} \tilde{G}^{\mu \alpha}) \ \right\} \, \ + \nonumber \\
&+& \frac{1}{M^2}\left\{ {D}^{\mu}\tta_X\left[ i(D^{\nu}H)^{\dagger}(c_1\tilde{F}^Y_{\mu\nu}+2c_2\tilde{F}^W_{\mu\nu})H+h.c.\right]\right.\ok
 &+&\left.\partial^mD_m \theta_X (d_1\mc{T}r(F^Y\tilde{F}^Y) +2d_2\mc{T}r(F^W\tilde{F}^W)) +
 d'_{ew} \partial^{\mu} D^{\nu} \theta_X {\rm Tr}(F_{\mu \rho}\tilde{F}^{\rho}_{\nu})  \nonumber
\right. \\
&+& \left. e_{ew}  D^{\mu} \theta_X {\rm Tr}(F_{\nu \rho} {\cal D}_{\mu} \tilde{F}^{\rho \nu})
+ e'_{ew} D_{\mu} \theta_X {\rm Tr}(F_{\alpha \nu} 
{\cal D}^{\nu} \tilde{F}^{\mu \alpha})\right\}\, \ ,   \label{gluon1} 
\end{eqnarray}
where ${\cal D}_{\mu} {G}_{\alpha \beta}$ denotes the gluon covariant derivative, in components
\begin{equation}
{\cal D}_{\mu} {G}_{\alpha \beta}^a \ = \ {\partial}_{\mu} {G}_{\alpha \beta}^a
+ g f^{abc} G_{\mu}^b G_{\alpha \beta}^c \ . 
\end{equation}
The last three terms in (\ref{gluon1}) refer to all electroweak gauge bosons.
  
{\bf iii)} The mediators do not form an anomaly-free set. It means that some low-energy fermions have to be charged in order to compensate the resulting anomaly. The induced dimension-six operators in this case are not gauge invariant, but include axionic
couplings and eventually GCS terms, schematically of the form
\begin{equation}
\mathcal{L} = \ C_{ij}^X \frac{a_X}{V} {\cal T}r (F^i {\tilde F}^j) +
E_{ij,k} \epsilon^{\mu \nu \rho \sigma} A_{\mu}^i  A_{\nu}^j
 F_{\rho \sigma}^k \ . \label{gcs}
\end{equation}
This case was studied from various perspectives in the past \cite{abdk,ferrara} and will not be considered anymore here. 

In all cases, there is potentially a kinetic mixing term \cite{keith}
\begin{equation}
\frac{\delta}{2} \ F^{\mu \nu}_X \  F_{\mu \nu}^Y \ . \label{mixing}
\end{equation}
Mediators generate at one-loop $\delta \sim \frac{g_X g'}{16 \pi^2} \sum_i X_i Y_i \ln \frac{\Lambda^2}{M_i^2}$, where $X_i,Y_i$ are the mediators charges to $U(1)_X$ and $U(1)_Y$, respectively. If $\delta$ has its natural one-loop value, then
its effects are more important than most of the ones we will discuss in what follows. This is the most plausible case and was investigated in many details within the last years. In what follows, we will place ourselves in the mostly `orthogonal'
case in which $\delta$ is small enough such that its effects are subleading compared
to the dimension-six  operators. This is the case, for example, if messengers are in complete representations of a non-abelian gauge group (GUT groups are of course the best such candidates), or if the mediators have no hypercharge.  

Then, at low energy, the mediators being integrated out  give rise to a new effective lagrangian
\begin{eqnarray}
{\cal L}_{eff} = {\cal L}_{1}(\psi^{\rm DM},Z'_{\mu}) + {\cal L}_{2}(A_{\mu}^{SM}) + {\cal L}_{mix}(Z'_{\mu}, A_{\mu}^{SM}) \ ,
\label{inv2}
\end{eqnarray}
where ${\cal L}_{2}$ and ${\cal L}_{1}$ represent the new effective operators
generated separately in the SM gauge sector and $Z'$ one, whereas in
 ${\cal L}_{mix}$ we collect all the induced terms mixing $Z'$ with the Standard Model. Notice that ${\cal L}_1$ also contains the DM particle (i.e. the lightest mediator) which is not integrated out.

\begin{figure}[htbp]
\begin{center}
\includegraphics[width=15cm]{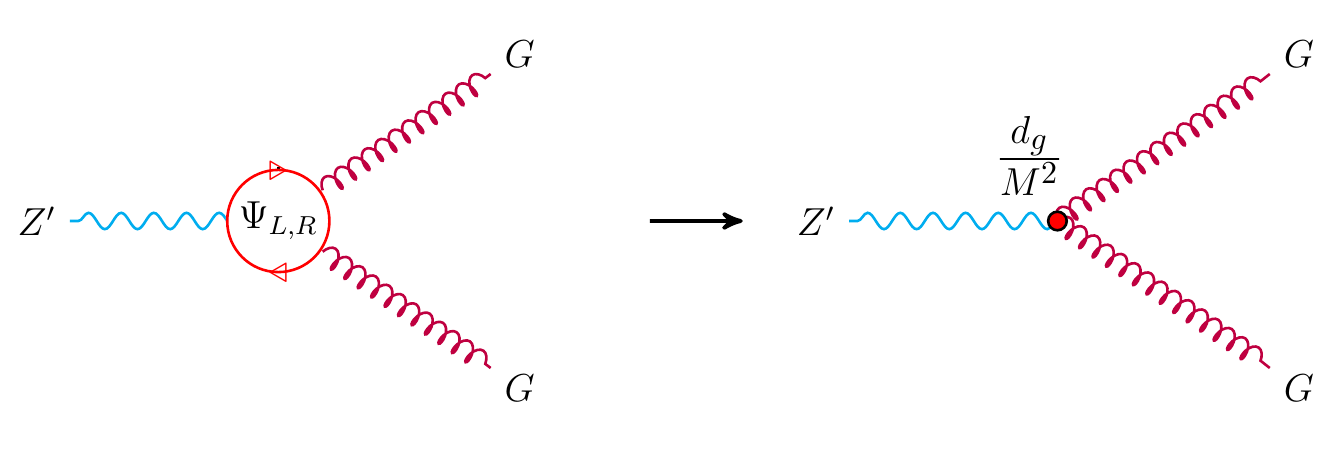}
\end{center}
\caption{\label{fig:GG1}{\em When heavy fermions are integrated out, they generate dimension-six effective operators of strength $d_g/M^2$.}}
\end{figure}

The mediators mass matrix has the symbolic form
\begin{equation}
M_{ij} \ = \ \lambda_{ij} V + h_{ij} v \ , \label{inv3}
\end{equation}
where $V$ is the vev breaking the $Z'$ gauge group $U(1)_X$ and $v$ is the electroweak vev. If the heavy Higgs $\phi$ has a charge $1$, then the renormalizable Yukawas (\ref{inv3}) exist provided
\begin{equation}
\lambda_{ij} \not=0 \text{ (and $h_{ij} =0$) } \quad {\rm if} \quad X_L^i-X_R^i = \pm 1~ , \quad
h_{ij} \not=0 \text{ (and $\lambda_{ij} =0$) }\quad {\rm if} \quad X_L^i-X_R^i = 0 \ . \label{inv4} 
\end{equation} 
Since none of our results in what follows depend on the assumption that the heavy fermions masses
arise through renormalizable interactions, in the rest of the paper we include the more
general case where these masses arise from arbitrary Yukawas of type
\[ \lambda_{ij} \Lambda (V/\Lambda)^{|X_L^i-X_R^j|} {\bar \Psi}_L^i \Psi_R^j + h.c\] where $\Lambda$ is an UV cut off, such that  $|X_L^i-X_R^j|> 1$ corresponds to
non-renormalizable interactions. 
For phenomenological applications, we consider here a model in which the dark matter is represented by the lightest stable fermion $\psi^{DM}$ charged under $Z'$ and uncharged under SM (the mass of dark matter will be denoted by $m_{\psi}$ in what follows). The {\em mediators} $\Psi_{L,R}$ are considered to be heavy enough so that they have not been discovered yet in colliders. They can be integrated out so that we have to deal with effective operators, including new parameters.
 At the one-loop perturbative level, mediators generate only $Z'$ couplings to the SM gauge fields and the SM Higgs as represented in Fig. \ref{fig:GG1} in the case of $Z'$ coupling to gluons. Indeed, in the absence of kinetic mixing, one-loop couplings to SM fermions can be generated only if there are Yukawa couplings mixing mediators with SM fermions. We forbid such couplings in what follows. One (clearly not unique) way of achieving this is  by defining a $Z_2$ parity, under which  all mediator fields are 
odd and all SM fields are even. 

In what follows we work in the unitary gauge where the axion is set to zero $\theta_X=0$. As usual, gauge invariance allows to work in any gauge. In the Appendix we discuss the issue of
gauge independence in more details.  

%%%%%%%%%%%%%%%%%%%%%%%%%%%%%%%%%%%%%%%%%%%%%%%%%%%%%%%%%%%%%%%
\subsection{Effective action from heavy fermion loops: coupling to gluons}

\begin{figure}[htbp]
\begin{center}
\includegraphics[width=8cm]{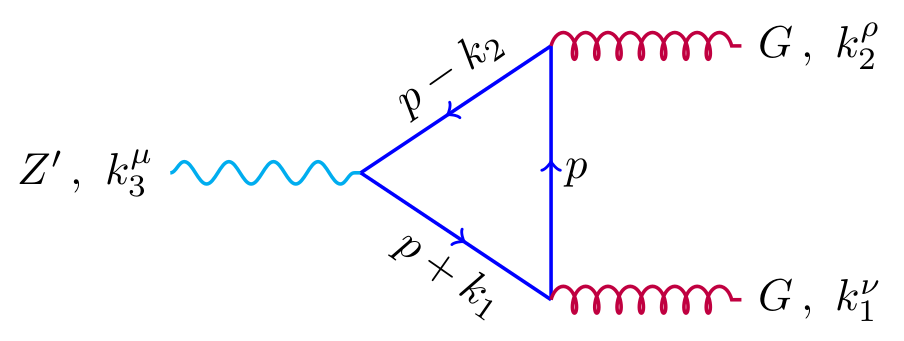}
\caption{\label{fig:triangle}\footnotesize{ Integration of heavy fermions in a triangle diagram.}}
\end{center}
\end{figure}

In the case of exact CP invariance that we restrict for simplicity, the three-point gauge boson amplitude can be generally be written
as \cite{abdk}
\begin{eqnarray}
&& \Gamma^{\mu \nu \rho} = \epsilon^{\mu \nu \rho \alpha} ( A_1 k_{1 \alpha}
+ A_2 k_{2 \alpha}) + \nonumber \\
&& \left[  \epsilon^{\mu \nu \alpha \beta} (B_1 k_1^{\rho} + B_2 k_2^{\rho} )
+ \epsilon^{\mu \rho \alpha \beta} (B_3 k_1^{\nu} + B_4 k_2^{\nu} ) \right] 
\ k_{1 \alpha} k_{2 \beta} \ , \label{g1} 
\end{eqnarray}
where $A_i,B_i$ are Lorentz-invariant functions of the external momenta $k_i$. 
The functions $A_i$ which encode the generalized Chern-Simon terms (GCS) \cite{abdk} are superficially logarithmically divergent, whereas the functions $B_i$
are UV finite. However, $A_i$ are determined in terms of $B_i$ by using the Ward identities, 
which in the case where the heavy fermions form an anomaly-free set, are given by 
\begin{eqnarray}
 k_1^{\nu} \Gamma_{\mu \nu \rho} \ &=& \ 0 \quad \rightarrow \quad A_2 = B_3 k_1^2 + B_4 k_1 k_2 \ , \nonumber \\
k_2^{\rho} \Gamma_{\mu \nu \rho} \ &=& \ 0  \quad \rightarrow \quad
A_1 = B_2 k_2^2 + B_1 k_1 k_2 \ , \nonumber \\
 - (k_1+ k_2)^{\mu} \Gamma_{\mu \nu \rho} \ &=& \   \ (A_1-A_2) \ \epsilon_{\nu \rho \alpha \beta} k_1^{\alpha} k_2^{\beta} \ \not = \ 0 \ . \label{g2} 
\end{eqnarray} 
The violation of the $Z'$ current conservation may seem surprising. It encodes actually the fact that one generates
dimension-six operators, for which gauge invariance is realized \`a la Stueckelberg and indeed in the Appendix B it will be shown explicitly that $A_1 \not = A_2$.
There are several contributions to $\Gamma^{\mu \nu \rho}$. The first is the triangle loop diagram with no chirality flip/mass insertions, given by
\begin{equation}
\Gamma_{\mu \nu \rho}^{(1)} =  \sum_i t_{iaa} 
\int \frac{d^4 p}{(2 \pi)^4} \ {\rm Tr} \left[ \frac{\slp + {\slk}_2}{(p+k_2)^2 - M_i^2}
\gamma_{\rho}  \frac{\slp }{p^2 - M_i^2} \gamma_{\nu}
 \frac{\slp - {\slk}_1}{(p-k_1)^2 - M_i^2} \gamma_{\mu} \gamma_5 \right] \ . \label{g4}  
\end{equation}
where $t_{iaa} = {\rm Tr} (X_i T^a T^a)$. As shown in the Appendix B by using Ward 
identities, computing this diagram is enough in order to find the full amplitude.   
The final result for the $Z'$ couplings and the details of the computation are described in the Appendix B.  After symmetrization among the two gluon legs, 
one finds
\begin{equation}
\Gamma_{\mu \nu \rho}^{\cal O} =  - \sum_i  \frac{i t_{iaa, L-R}}{12 \pi^2 M_i^2} 
\{ [ 2 ( k_1 +  k_2)_{\mu} \epsilon_{\nu \rho \alpha \beta} - k_{1 \rho} \epsilon_{\mu \nu \alpha \beta} - k_{2 \nu} \epsilon_{\rho \mu \alpha \beta} ] k_1^{\alpha} k_2^{\beta} 
+ \epsilon_{\mu \nu \rho \alpha} k_1 k_2 (k_2-k_1)^{\alpha} \} \ ,   
\label{g6}
\end{equation}
where $t_{iaa, L-R} = {\rm Tr} ((X_L-X_R) T^a T^a)_i$. 
The corresponding dimension-six operator for the triangle diagram represented in Fig. \ref{fig:triangle} is then
\begin{equation}
{\cal O} \ = \ \frac{g_3^2}{24 \pi^2 } \sum_i \ {\rm Tr} \left(\frac{(X_L-X_R) T_a T_a}{M^2}\right)_i
\ \left[ \partial^{\mu} D_{\mu} \theta_X {\cal T}r (G {\tilde G}) - 2
D_{\mu} \theta_X {\rm Tr}(G_{\alpha \nu} {\cal D}^{\nu} \tilde{G}^{\mu \alpha})  \right] \ , 
\label{g5}
\end{equation}
where $g_3$ is the QCD strong coupling. 

On the other hand, by using the identities (\ref{identity}) in Appendix C, it can be shown that the antisymmetric part of the amplitude in the gluonic legs is zero, which is consistent with the fact that there is no possible dimension-six operator mixing $Z'$ to gluons, that is antisymmetric in the gluon fields. As a byproduct, we also find that the heavy mediators we are considering do not induce operators of the type 
\begin{equation}
\frac{1}{M^2} {\rm Tr} ( G_{\mu \nu} [ G^{\nu \lambda}, {\tilde G}_{\lambda}^{\mu} ] ) 
\  , \label{g7}
\end{equation}
that are completely antisymmetric in the three gluon fields \label{g5}, and similar
operators for electroweak gauge fields.   
This means that there are no constraints from purely SM dimension-six operators
induced in this setup and all the phenomenological constraints come from the mixing
of $Z'$ with SM fields. 
  
%%%%%%%%%%%%%%%%%%%%%%%%%%%%%%%%%%%%%%%%%%%%%%%%%%%%%%%%%%%%%%%%%%%%%%%%%%%
\subsection{``Anomalous" $Z'$}

Until now we have made the important assumption that no SM fermion is charged under $Z'$ and the
only couplings arise through gauge-invariant higher-dimensional operators generated by integrating out heavy fermions forming an anomaly-free set. A more subtle option,
in the spirit of \cite{abdk,wells,dmpr,tait} is to integrate-out a set of heavy fermions which
do contribute to gauge anomalies. In this case there are non-decoupling effects leading to
axionic couplings and eventually generalized Chern-Simons terms. Let us consider two simple examples in order to exemplify the main points.   

{\bf i) Example with no colour anomalies}
$$
\begin{array}{|c|c|c|c|}
\hline
{\rm \ Field} \quad &  \quad \ Q_3^L \quad & \quad t_{R} \quad & \quad b_{R}  \\
\hline
 Z' {\rm charge} \quad & \quad 1 \quad & \quad  1 \quad & \quad 1  \\
\hline
\end{array}
$$ \\
In this case, after defining the anomaly coefficients 
$C_a = {\rm Tr} (X T_a^2)_{L-R}$ and $C_X = {\rm Tr} (X^2 Y)_{L-R}$, 
the low-energy effective theory has the following mixed anomalies:
\begin{eqnarray}
&& U(1)_X SU(3)^2 \quad : \quad C_3 \ = \  \frac{1}{2} \times (2-1-1) = 0 \ , \nonumber \\
&& U(1)_X SU(2)^2 \quad : \quad C_2 \ = \ \frac{1}{2} \times 3  \ , \nonumber \\
&& U(1)_X U(1)_Y^2 \quad : \quad C_1 \ = \ 6 \times \frac{1}{9} - 3 \times (\frac{16}{9} + \frac{4}{9}) = -6 \  , \nonumber \\
&& U(1)_X^2 U(1)_Y \quad : \quad C_X \ = \ 6 \times \frac{1}{3} - 3 \times \frac{4}{3} + 3 \times \frac{2}{3} = 0 \  .
 \label{an1}
\end{eqnarray}

{\bf i) Example with colour anomalies}
$$
\begin{array}{|c|c|c|c|}
\hline
{\rm \ Field} \quad &  \quad \ Q_3^L \quad & \quad t_{R} \quad & \quad b_{R}  \\
\hline
 Z' {\rm charge} \quad & \quad 1 \quad & \quad  1 \quad & \quad 0  \\
\hline
\end{array}
$$ \\

In this case, the low-energy effective theory has the following anomalies:
\begin{eqnarray}
&& U(1)_X SU(3)^2 \quad : \quad  C_3 \ = \ \frac{1}{2} \times (2-1) = \frac{1}{2} \ , \nonumber \\
&& U(1)_X SU(2)^2 \quad : \quad  C_2 \ = \ \frac{1}{2} \times 3  \ , \nonumber \\
&& U(1)_X U(1)_Y^2 \quad : \quad  C_1 \ = \ 6 \times \frac{1}{9} - 3 \times \frac{16}{9} = - \frac{14}{3} \  , \nonumber \\
&& U(1)_X^2 U(1)_Y \quad : \quad C_X \ = \ 6 \times \frac{1}{3} - 3 \times 
\frac{4}{3} = -2 \ .  
 \label{an2}
\end{eqnarray}

In such examples, the heavy-fermion spectrum has to exactly cancel the low-energy gauge anomalies.
In the decoupling limit there is an axionic coupling with a coefficient exactly determined
by the low-energy induced anomalies
\begin{equation}
{\cal L}_{\rm ax} \ = \ \frac{a_X (x)}{16 \pi^2 V} 
\left[ \sum_a  (C_a  g_a^2
\ \epsilon^{\mu \nu \rho \sigma} F_{\mu \nu}^a F_{\rho \sigma}^a )
\ + \ C_X \ g_X g' \epsilon^{\mu \nu \rho \sigma} F_{\mu \nu}^X F_{\rho \sigma}^Y
\right] \  . \label{an3}
\end{equation}
As shown in the Appendix B, we can also capture the effect of these axionic couplings in the unitary
gauge, where the axionic effect is encoded in the particular high-energy behaviour of the
anomalous three gauge boson amplitude with light fermions in the loop. This is strictly
speaking true in the large (infinite) mass limit of heavy fermions. For finite mass, there
are corrections and the low-energy description in the unitary gauge with three-gauge anomalous couplings is corrected by finite mass effects.

%%%%%%%%%%%%%%%%%%%%%%%%%%%%%%%%%%%%%%%%%%%%%%%%%%%%%%%%%%%%%%%%%%%%%%%%%%%%%%%%%%%%%%%
\section{Dark Matter Annihilation to gluons}

We start by first discussing the $Z'$ couplings to gluons. 
The CP and gauge invariant dimension-six operators coupling $Z'$ and the gluons are given by
\begin{eqnarray}
 \mathcal{L}_{\text{CP even}}&=&\frac{1}{M^2}\left\{ d_g\partial^{\mu} D_{\mu} \theta_X {\cal T}r(G\tilde{G}) + 
d'_g \partial^{\mu} D^{\nu} \theta_X {\rm Tr}(G_{\mu \rho}\tilde{G}^{\rho}_{\nu})  \nonumber
\right. \\
&+& \left. e_g  D^{\mu} \theta_X {\rm Tr}(G_{\nu \rho} {\cal D}_{\mu} \tilde{G}^{\rho \nu})
+ e'_g D_{\mu} \theta_X {\rm Tr}(G_{\alpha \nu} 
{\cal D}^{\nu} \tilde{G}^{\mu \alpha}) \ \right\} \, \ .  \label{gluon01} 
\end{eqnarray}
Due to the fact that at one-loop only
the operators with coeff. $d_g$ and $e'_g$ are generated and only the first one contributes to the amplitude with on-shell gluons, we consider only $d_g$ in what follows and disregard the effects of the other operators in (\ref{gluon01}).  

\begin{figure}[htbp]
\begin{center}
\includegraphics[width=12cm]{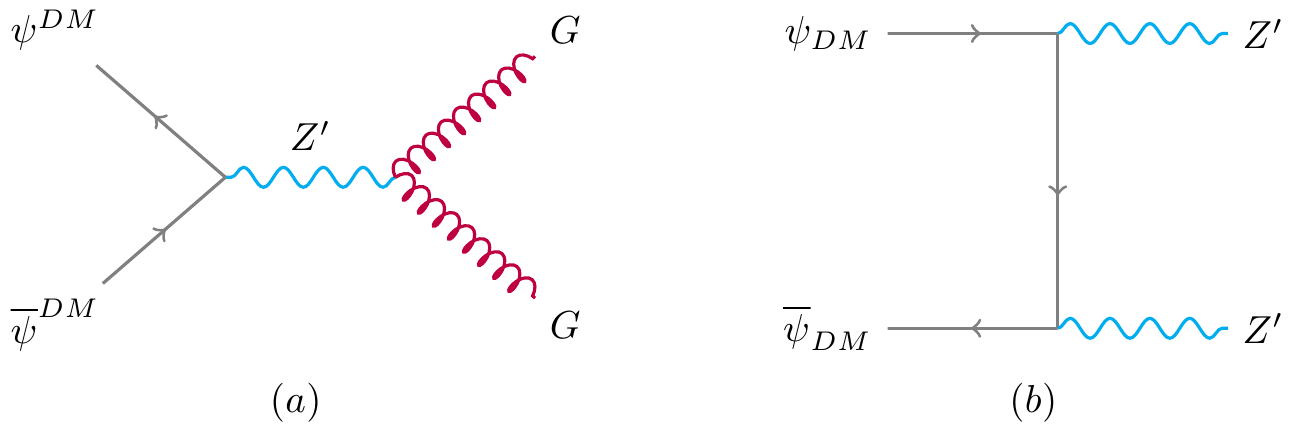}
\caption{\label{fig:GG}{\em Gluon pair production via two different processes, that are the s-channel $(a)$ and the t-channel $Z'$ pair production $(b)$, that decay subsequently into two gluons each.}}
\end{center}
\end{figure}

The {\em dark matter} couples minimally to the $Z'$ boson as:
 \begin{equation}
  \bar{\psi}^{DM}_L\half{g_X}X^{DM}_L\gamma^{\mu}Z'_{\mu}\psi^{DM}_L + \bar{\psi}^{DM}_R\half{g_X}X^{DM}_R\gamma^{\mu}Z'_{\mu}\psi^{DM}_R\,,
 \end{equation}
 which provides us two ways of annihilating dark matter into gluons. The first one is an {\em s-channel} production of a $Z'$ boson decaying into a pair of gluons. The second one is a {\em t-channel} process, leading to two $Z'$ bosons, which will mostly decay into gluons. The associated Feynman diagrams are presented in Fig.\ref{fig:GG}.
\newline

In the unitary gauge, the $Z'$-gluon-gluon vertex coming from the operator  $d_g$ is
\begin{equation}
 {\frac{d_g}{M^2}}\left\{g_X\partial^mZ'_m\epsilon^{\mu\nu\rho\sigma}
 \partial_{\mu}G^A_{\nu}\partial_{\rho}G^A_{\sigma}\right\}\, ,
\end{equation} 
where the coeff. $d_g$ was redefined compared to (\ref{gluon01}) in a convenient way
for our purposes. 

The propagator of the vector boson $Z'$ in the unitary gauge is
\begin{equation}
 \Delta(q) = - i \ \frac{\eta_{\m\n} - \frac{q_{\m}q_{\n}}{M_{Z'}^2} }{q^2-M_{Z'}^2+i M_{Z'}\Gamma(Z')}\, \ ,
\end{equation}

For dark matter fermions of mass smaller than $M_{Z'}/2$, the main contribution to 
the $Z'$ width $\Gamma(Z')$ is $\Gamma(Z'\rightarrow \psi^{DM} \overline{\psi^{DM}})$, 
which is computed to be
\begin{equation}
 \Gamma(Z') = \frac{g_X^2}{384\pi M_{Z'}^2} \left[(X_L^2+X_R^2)M_{Z'}^2 - (X_L^2+X_R^2-6X_RX_L)m_{\psi}^2 \right] \sqrt{M_{Z'}^2-4m_{\psi}^2}\, \ .
\end{equation} 

For heavier masses of dark matter, one has to consider the $Z'$ decay width into gluons and $SU(2)$ gauge bosons. However, it can be readily checked that 
the detailed values of these widths do not influence much the results in what follows\footnote{Indeed, we will see in what follows that the cross section of dark
 matter annihilation into gluons is suppressed for an invariant mass $\sqrt{s}$ approaching $M_{Z'}$, as a consequence of the Landau-Yang theorem 
 \cite{landauyang}. In the non-relativistic approximation, this happens in the energy region closed to  $s\simeq 4 m_{\psi}^2 + m_{\psi}^2v_{rel}^2 \geqslant 4 m_{\psi}^2$. 
 The suppression therefore occurs for a mass $m_{\psi}$ significantly lower than $M_{Z'}/2$, where the decay width is essentially that of decay into two dark matter particles.}.
%%%%%%%%%%%%%%%%%%%%%%%%%%%%%%%%%%%%%%%%%%%%%%%%%%%%%%%%%%%%%%%%%

%==============================  S-CHANNEL ======================================================
\subsection{ The {\em s-}channel dark matter-gluons cross-section}
\subsubsection{Vector-coupling case}
\label{Sec:vector}

\noindent
In the case of a vector-like coupling of DM fermion to $Z'$ boson, one obtains the interaction lagrangian
\begin{equation}
{\cal L}_{int} = \bar{\psi}^{DM}\half{g_X}X^{DM}\gamma^{\mu}Z'_{\mu}\psi^{DM},
\quad {\rm where} \quad \text{~~~~~~~}X^{DM}\equiv X^{DM}_R=X^{DM}_L\,.
\end{equation}

\noindent
Now we can perform the tree-level diagram cross section. We find that the amplitude vanishes $\mathcal{M} =0$
and therefore the $d_g$-term does not contribute to the final cross section at all. The reason is that, due to the effective coupling of the form 
$ d_g \partial^m Z'_m  {\cal T}r(G\tilde{G})$, the vertex $Z' \psi^{DM} \psi^{DM}$
gets multiplied by the virtual momentum and is of the form
\begin{equation}
q^{\mu} {\bar v} (p_2) \gamma_{\mu} u (p_1) =  {\bar v} (p_2) 
({\slp}_2 +{\slp}_1 ) u (p_1) = 0 \ ,
\end{equation}
after using Dirac equation for the spinors describing the wavefunctions of the dark matter fermions. 
%%%%%%%%%%%%%%%%%%%%%%%%%%%%%%%%%%%%%%%%%%%%%%%%%%%%%%%%%%%%%%%%%%%%%%%%%%%%%%%%%%%
\subsubsection{Axial-vector couplings case}
In the general case we get also an axial-vector coupling in addition to the vector one
\begin{equation}
{\cal L}_{int} =  \half{g_X}\left(\frac{X_R^{DM}+X_L^{DM}}{2}\right)\bar{\psi}^{DM}\gamma^{\mu}Z'_{\mu}\psi^{DM} + \half{g_X}\left(\frac{X_R^{DM}-X_L^{DM}}{2}\right)\bar{\psi}^{DM}\gamma^{\mu}\gamma_5Z'_{\mu}\psi^{DM}\,.
\end{equation}

One then gets, as far as the annihilation of dark matter into a gluon pair is concerned, the total cross section
\begin{eqnarray}
 &&\sigma_{s-channel}(\psi^{DM} \psi^{DM} \rightarrow GG)={\frac{d_g^2}{M^4}}\frac{(-4E^2+M_{Z'}^2)^2}{(-4E^2+M_{Z'}^2)^2+ M_{Z'}^2\Gamma^2(Z')}\frac{ E^5 g_X^4 m_{\psi}^2 ({X_L}-{X_R})^2}{\pi  M_{Z'}^4 \sqrt{E^2-m_{\psi}^2}}\,.\ok
\end{eqnarray}

The cross section is suppressed for energies of order $M_{Z'}/2$ due to the Landau-Yang theorem.  There is also a  helicity suppression for light dark matter case, that can be easily understood
by  writing the vertex $Z' \psi^{DM} \psi^{DM}$ in this case
\begin{equation}
({X_L}-{X_R}) q^{\mu} {\bar v} (p_2) \gamma_{\mu} \gamma_5 u (p_1) =  ({X_L}-{X_R})
{\bar v} (p_2)  ({\slp}_2 \gamma_5 - \gamma_5 {\slp}_1 ) u (p_1) = 
 -2 m_{\psi}({X_L}-{X_R}) {\bar v} (p_2)  \gamma_5 u (p_1),
\end{equation}
after using Dirac equation.  

This finally gives in the non-relativistic approximation $s\simeq4m_{\psi}^2 + m_{\psi}^2v_{rel}^2 \Leftrightarrow E \simeq m_{\psi}\sqrt{1+\frac{v_{rel}^2}{4}}$, with $v_{rel}$ 
being the relative velocity between the two colliding dark matter fermions, the total cross section
\begin{equation}
\langle\sigma v\rangle_{s-channel} \simeq {\frac{d_g^2}{M^4}}\frac{g_X^4m_{\psi}^6 ({X_L}-{X_R})^2}{\pi M_{Z'}^4}\left\{\frac{2 \left(M_{Z'}^2-4 m_{\psi}^2\right)^2 }{ \left(M_{Z'}^2\Gamma^2(Z')+\left(M_{Z'}^2-4 m_{\psi}^2\right)^2\right)}\right\}+\mathcal{O}\left(v^2\right)
%&\equiv& \blue{\frac{d_g^2}{M^4}} \langle\Sigma v\rangle_{s-channel}\,.
\label{Eq:schannel}
\end{equation}

%====================================  T-CHANNEL  =================================================

\subsection{ The {\em t}-channel dark-matter decay}

As mentioned earlier, we also have to consider a t-channel process, producing pairs of $Z'$ bosons in dark matter annihilation for $Z'$ mass below the dark matter mass. 
Considering that the only non vanishing coupling is the one in $d_g$, each $Z'$ will decay into gluons; this process will then produce gluons in the final state.
 After expanding in powers of $v^2$, the cross-section in this case can be expressed as:
\begin{eqnarray}
&& \langle\sigma v\rangle_{t-channel}=\frac{g_X^4 \sqrt{m_{\psi}^2-M_{Z'}^2}}{128 \pi ^2 m_{\psi} M_{Z'}^2 \left(2 m_{\psi}^2-M_{Z'}^2\right)^2}\left(2 m_{\psi}^4 X_L^4-4 m_{\psi}^4 X_L^2 X_R^2+2 m_{\psi}^4 X_R^4-3 m_{\psi}^2 M_{Z'}^2 X_L^4\right.\ok
&&+ \left.10 m_{\psi}^2 M_{Z'}^2 X_L^2 X_R^2-3 m_{\psi}^2 M_{Z'}^2 X_R^4+M_{Z'}^4 X_L^4-6 M_{Z'}^4 X_L^2 X_R^2+M_{Z'}^4 X_R^4\right)+\mathcal{O}\left(v^2\right)\,.\ok\ok
 \label{Eq:tchannel}
\end{eqnarray}

%====================================================================================================
%====================================  CONSTRAINTS    =================================================
%====================================================================================================

\section{Experimental constraints}

\noindent
A $Z'GG$ coupling can be tested in several laboratories, from direct detection experiments to indirect detection, relic abundance or LHC searches.
We present in the following the constraints obtained from these different  searches, before summarizing all of them at the end of the section.
The reader can also find a nice recent complementary analysis of gluonic effective couplings to dark matter in \cite{Chu:2012qy}.

%===================================   RELIC ABUNDANCE   =============================================

\subsection{Relic abundance}

\noindent
Recently, PLANCK collaboration released its latest results concerning the composition of the Universe \cite{Ade:2013zuv}.
It confirms the results of WMAP experiment \cite{WMAP} obtaining for the relic abundance of non--baryonic matter
$\Omega h^2 =0.1199 \pm 0.0027$ at 68\% of CL. With such a level of precision, it is interesting to know what is the 
effective scale $M$ which is able to produce sufficient dark matter from the thermal bath to respect the previous PLANCK/WMAP results.
Depending on the spectrum, two annihilation processes allow the dark matter candidate to keep thermal equlibrium
with the standard model particles of the plasma: the $s$--channel exchange of a $Z'$ (Eq.\ref{Eq:schannel}), and
 the $t$-channel production of the $Z'$ (Eq.\ref{Eq:tchannel}), as long as this channel is kinematically open. 
 
 \noindent
 Concerning the numerical analysis, we solved the Boltzmann equations by developing a code and adapting the public software 
 MicrOMEGAs \cite{Belanger:2010gh} to our model.
We then extracted the relic abundance and checked that our analytical solutions (\ref{Eq:schannel}-\ref{Eq:tchannel}) gives
 similar results to the numerical procedure\footnote{Mainly because the dominant annihilations are dominated
 by s--wave processes and the solution $\langle \sigma v \rangle \simeq 3 \times10^{-26} ~\mathrm{cm^3 s^{-1}}
 \simeq 2.5 \times 10^{-9} ~\mathrm{GeV^{-2}}$ gives reasonable
 good approximations to the full Boltzmann system of equations.} at a level of 20 to 30\%.
 We noticed in section \ref{Sec:vector} that the coupling of the dark matter should be axial, as the vectorial part of the current
 coupling to $Z'_\mu$ does not gives any contribution to the process $\psi^{DM} \psi^{DM} \rightarrow Z' \rightarrow GG$.
  For simplicity, we will  set charges $X_R=1, X_L=2~\Rightarrow~  |X_R-X_L|=1$. Our results
for a  different set of charges are modified in a straightforward way. To keep our results as conservative as possible,
 we plotted the WMAP limits $ 0.087 < \Omega h^2 < 0.138$ at 5$\sigma$.

%%%%%%%%%%%%%%%%%%%%%%%%%%%%%%%%%%%%%%%%%%%%%%%%%%%%%%%%%%%%%%%%%%%
 \begin{figure}
% $T_{RH}$\includegraphics[width=0.5\textwidth,angle=0]{Reheat.pdf}
%\includegraphics[width=0.5\textwidth,angle=0]{YvsX_vecscalDM.pdf}
\begin{tabular}{c c}
\includegraphics[width=0.5\textwidth,angle=0]{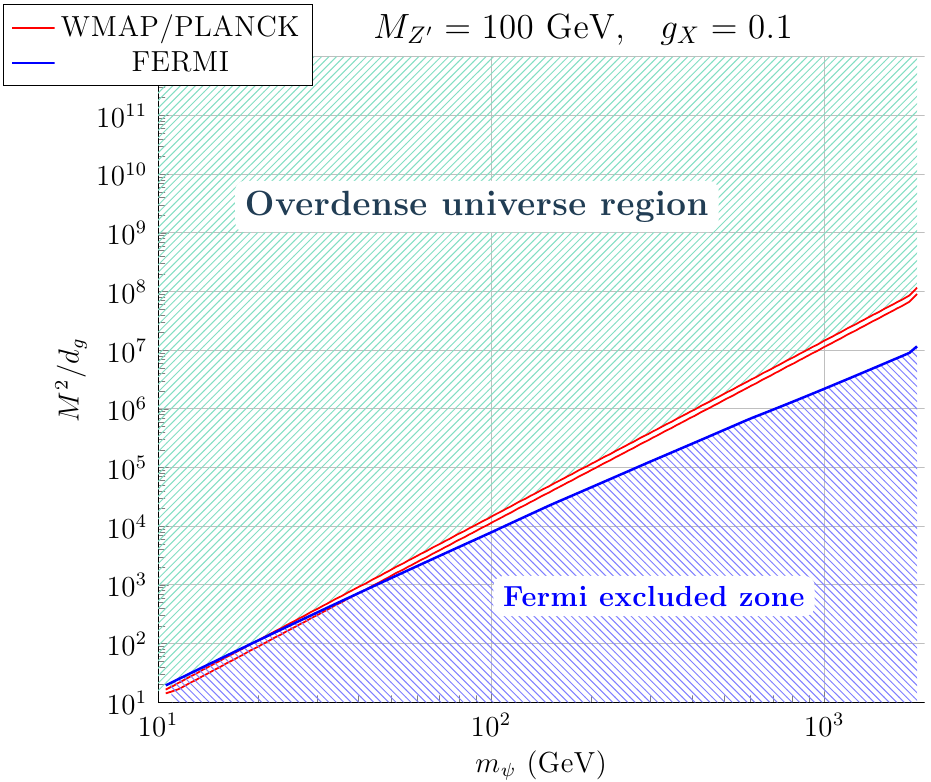} & \includegraphics[width=0.5\textwidth,angle=0]{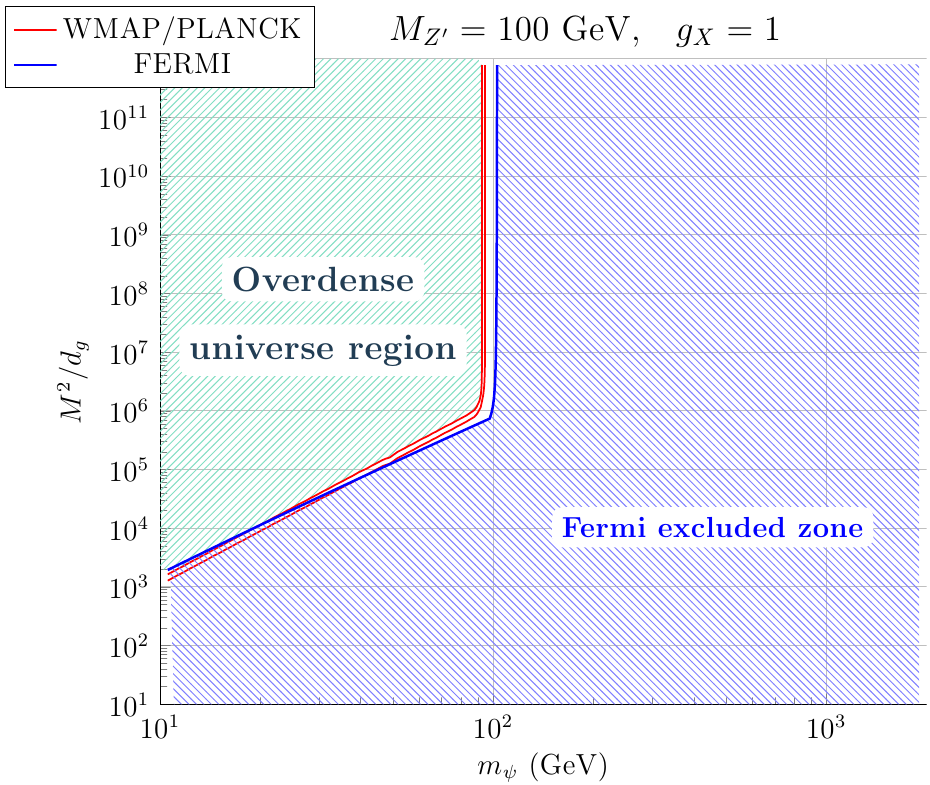} \\
\includegraphics[width=0.5\textwidth,angle=0]{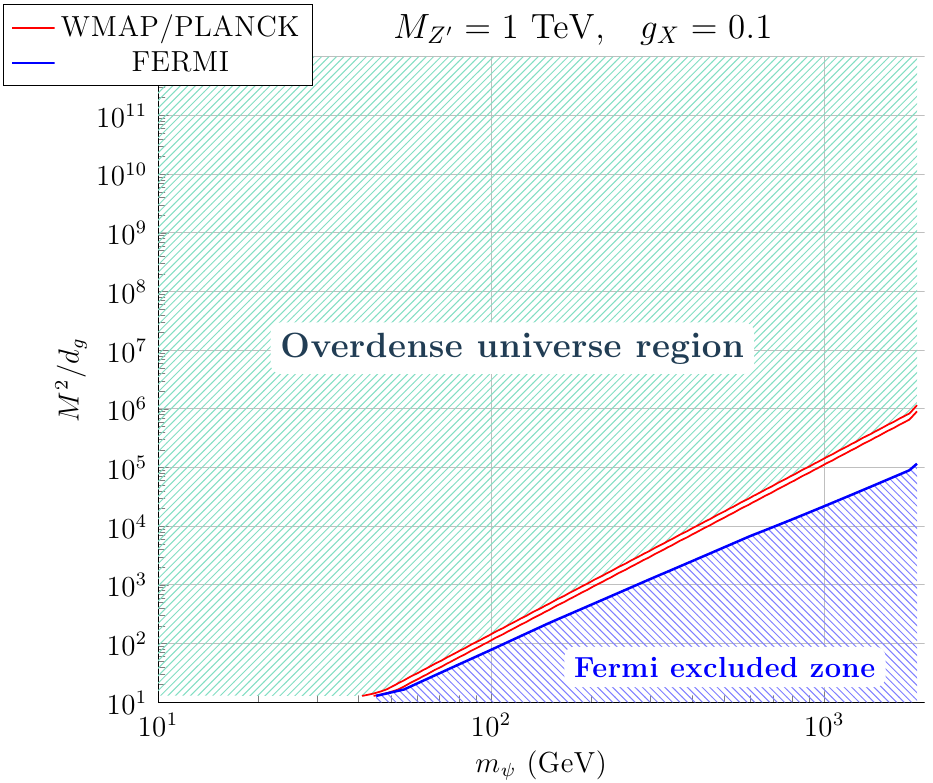} & \includegraphics[width=0.5\textwidth,angle=0]{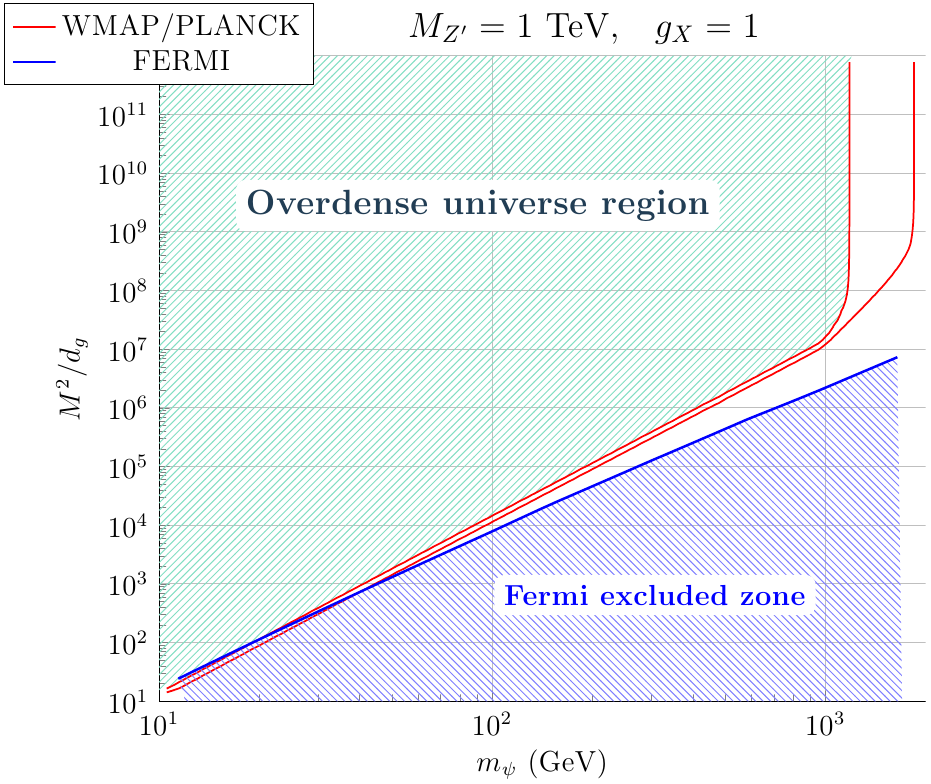}
\end{tabular}
\caption{\label{Fig:WMAP}\footnotesize{Constraints from WMAP/PLANCK (red line) and FERMI dSphs galaxies (blue line) in
the ($\frac{M^2}{d_g}, m_\psi$) plane for different values of $g_X$ (0.1 on the left  and 1 on the right), $M_{Z'}=100$ GeV (up) and $M_{Z'}=1$ TeV
(down). See the text for more details.}}
\end{figure} 
%%%%%%%%%%%%%%%%%%%%%%%%%%%%%%%%%%%%%%%%%%%%%%%%%%%%%%%%%%%%%%%%%%%%%

\noindent
We show in Fig. \ref{Fig:WMAP} the parameter space allowed in the plane ($\frac{M^2}{d_g}, m_\psi$) for different values of $M_{Z'}$ and
$g_X$. Points $above$ the red lines region would lead to an overpopulation of dark matter whereas points lying $below$ the red lines
would require additional dark matter candidates to respect PLANCK/WMAP constraints.
We can notice several, interesting features from these results. First of all, we observe that as soon as the $Z'Z'$ final state
is kinematically allowed ($m_\psi > M_{Z'}$) this annihilation channel is the dominant one as soon as $g_X$ is sufficiently large
(we checked that this happens for $g_X \gtrsim 0.3$) and mainly independent on the dark matter mass. This is easy to understand after an inspection of  
Eq.(\ref{Eq:tchannel}). Indeed, in the limit $m_\psi \gg M_{Z'}$, one obtains $\langle \sigma v \rangle_{Z'Z'} \simeq \frac{9 g_X^4}{256 \pi^2 M_{Z'}^2}$. 
In other words, once 

\begin{equation}
\frac{9~g_X^4}{256 \pi^2 M_{Z'}^2} \gtrsim 2.5 \times 10^{-9} ~\mathrm{GeV^{-2}} ~~\rightarrow ~~ g_X \gtrsim 3 \times 10^{-2} \sqrt{\frac{M_{Z'}}{{\rm GeV}}}  \ , 
\label{Eq:limit}
\end{equation}

\noindent
then the $t$-channel process $\psi^{DM} \psi^{DM} \rightarrow Z' Z'$ dominates the annihilation and forbids the dark matter to
overpopulate of the Universe ($\Omega_\psi h^2 \lesssim 0.138$). This corresponds to $g_X \simeq 0.3$ for $M_{Z'}=100$ GeV
and $g_X \simeq 1$ for $M_{Z'}=1$ TeV, which fits pretty accurately the numerical results we obtained. 
This limit also explains why the region allowed by PLANCK/WMAP is larger for $M_{Z'}= 1$ TeV: the value $g_X=1$ is
at the border limit for the $t-$channel to dominate. From Eq.(\ref{Eq:limit}) we also understand why the $Z'Z'$ final state, even
if kinematically allowed, has no influence on the limits set by the relic abundance for $g_X=0.1$: the coupling is too small
to give sufficient annihilation products. The dominant process is then the $s-$channel $Z'$ exchange ($\simeq 15\%$ of $Z'Z'$ final
state for $g_X=0.1$ and $M_{Z'}=1$ TeV.).

\noindent
A different choice for the charges $X_L$ and $X_R$ has a straightforward influence on this result since it will change an overall factor in Eq. (\ref{Eq:limit}). As an example, taking $X_R=5$ and $X_L=6$ will give

\begin{equation}
\langle \sigma v \rangle_{Z'Z'} \simeq\frac{121~g_X^4}{256 \pi^2 M_{Z'}^2} \gtrsim 2.5 \times 10^{-9} ~\mathrm{GeV^{-2}} ~~\rightarrow ~~ g_X \gtrsim 1.5  \times 10^{-2} \sqrt{\frac{M_{Z'}}{{\rm GeV}}}  \ , 
\label{Eq:limit2}
\end{equation}
implying that the t-channel will become dominant for $g_X \simeq 0.1$ for $M_{Z'}=100$ GeV
and $g_X \simeq 0.4$ for $M_{Z'}=1$ TeV. The parameter space will then be slightly enlarged.

\noindent
We also notice in Fig. \ref{Fig:WMAP} that the region of the parameter space respecting WMAP/PLANCK data with a dominant $s-$channel annihilation
seems linear (in logarithmic scale). This can be easily understood; indeed, after a glance at Eq.(\ref{Eq:schannel}), one obtains \footnote{Neglecting
the tiny region around the pole $M_{Z'} = 2 m_\psi$.}

\begin{equation}
\langle \sigma v \rangle \simeq \frac{d_g^2}{M^4} \frac{2 g_X^4}{\pi} \frac{m^6_\psi}{M_{Z'}^4} ~~~( {\rm for} \ M_{Z'} \gg m_\psi ~~ \mathrm{or} ~~ M_{Z'} \ll m_\psi) 
\ , 
\end{equation}

\noindent
which imply for constant $\langle\sigma v \rangle $,

\begin{equation}
\log \left( \frac{M^2}{d_g} \right) = 3 \log m_\psi + \mathrm{const} \ , 
\end{equation}
\noindent which is exactly the behavior we observe in Fig.\ref{Fig:WMAP}.

\subsection{Indirect detection of dark matter}

\noindent
Other astrophysical constraints arise from the diffuse gamma ray produced by the dark matter annihilation in the center of Milky Way \cite{Hooper:2012sr},
the galactic halo \cite{Ackermann:2012rg}, the dwarf spheroidal galaxies \cite{DrlicaWagner:2012ry} or the radio observation of 
nearby galaxies like M31 \cite{Egorov:2013exa}. Even if the authors of \cite{Egorov:2013exa} claimed that their limits
``{\it exceed the best up-to-day known constraints from Fermi gamma observations}", the dependence on magnetic fields profiles
and charged particles propagation in M31 medium brings some uncertainties difficult to evaluate. The same remark is valid for the
galactic center study \cite{Hooper:2012sr} where the region of the sky and the cut made to analyze the data depends strongly on
the dark matter halo profile in play to maximize the signal/background ratio. We will then consider the more reliable constraints obtained
by the observation of dwarf galaxies by the FERMI telescope \cite{DrlicaWagner:2012ry}. These galaxies being mainly composed of dark matter,
the background is naturally minimized.

\noindent
We show the result of our analysis in Fig.\ref{Fig:WMAP} where the points $below$ and on the $right$ of the blue lines are excluded
by FERMI observations. As expected, the region below $m_\psi \lesssim 40-50$ GeV (where the curves from FERMI and WMAP/PLANCK cross) is in tension with FERMI limit, 
as hadronic final states
are the more restricted by FERMI analysis\footnote{Notice however that FERMI considers in their analysis the $Z'$ decays into quarks, whereas in our case it decays into gluons.}, which seems to exclude any thermal relics below this dark matter mass. When the $Z'Z'$ final state is allowed,
the annihilation cross section $\psi \psi \rightarrow Z' Z'$ is so large that is is almost automatically excluded by FERMI data.

%%%%%%%%%%%%%%%%%%%%%%%%%%%%%%%%%%%%%%%%%%%%%%%%%%%%%%%%%%%%%%%%%%%%%%%%%%%%%%%%%%%
\subsection{Direct Detection}

For direct detection purposes, one can integrate out the $Z'$ gauge boson
and write the corresponding dimension-eight operator connecting the dark matter with the gluons.
One gets
\begin{equation}
\frac{d_g}{M^2 M_{Z'}^2} \ \bar{\psi}^{DM}\gamma^{\mu} \left( \frac{X_R+X_L}{2}+\frac{X_R-X_L}{2} \gamma_5 \right) \psi^{DM}
{\cal T}r \ \partial_{\mu} ( G {\tilde G}) \ .  
\end{equation}
By using the observed CP invariance of the strong interactions, we find that the only non-vanishing relevant gluonic matrix element  we can write  between an initial and a final nucleon state is
$ \langle N(p) | {\rm Tr} \ G_{\mu}^{\nu} {\tilde G}_{\nu}^{\lambda} | N(p') \rangle = 
A \epsilon_{\mu}^{\lambda \alpha \beta} p_{\alpha} p'_{\beta}$, where $A$ is a Lorentz invariant. As a consequence, 
\begin{equation}
\langle N(p) | {\cal T}r \ \partial_{\mu} ( G {\tilde G}) | N(p') \rangle = 0 \ .  
\end{equation}
There are therefore no constraints on this operator from direct detection experiments. 
%%%%%%%%%%%%%%%%%%%%%%%%%%%%%%%%%%%%%%%%%%%%%%%%%%%%%%%%%%%%%%%%%%%%%%%%%%

%========================================================================================================================
%=====================================   LHC bounds  =======================================================================
%========================================================================================================================

\subsection{LHC analysis through mono-jets}

\begin{figure}[htbp]
\begin{center} 
\includegraphics[width=0.3\textwidth]{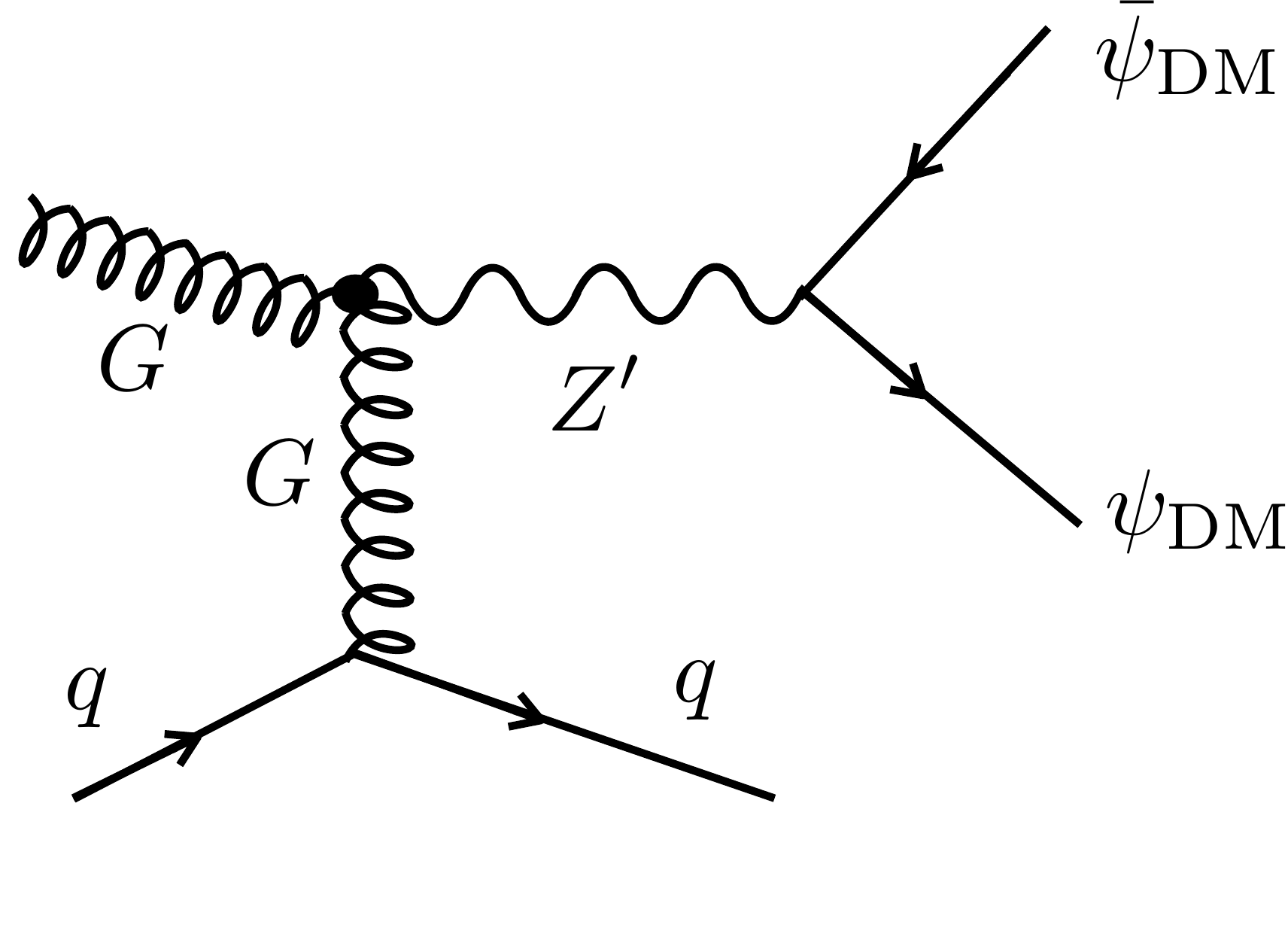}
\includegraphics[width=0.3\textwidth]{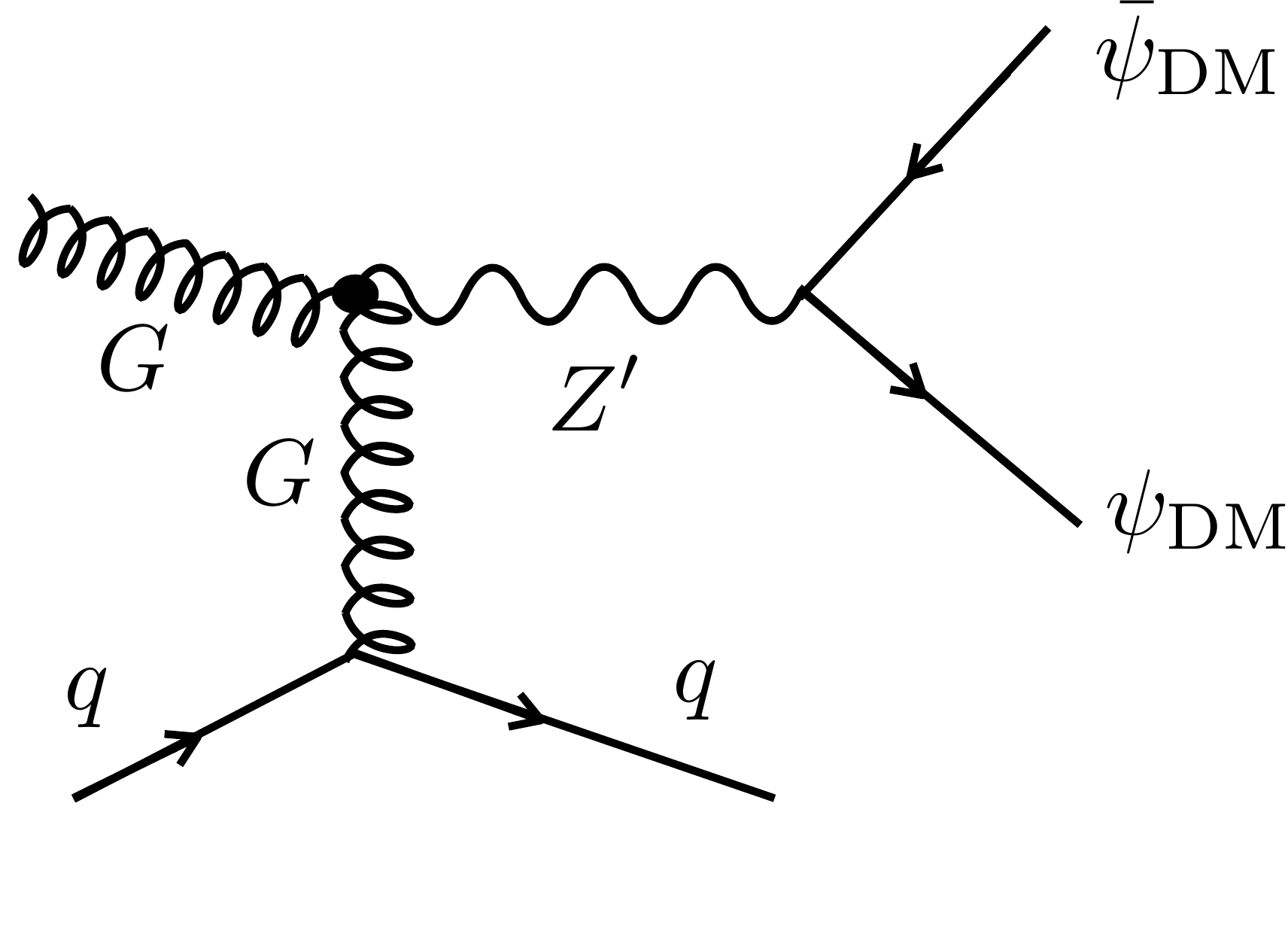} \\
\includegraphics[width=0.3\textwidth]{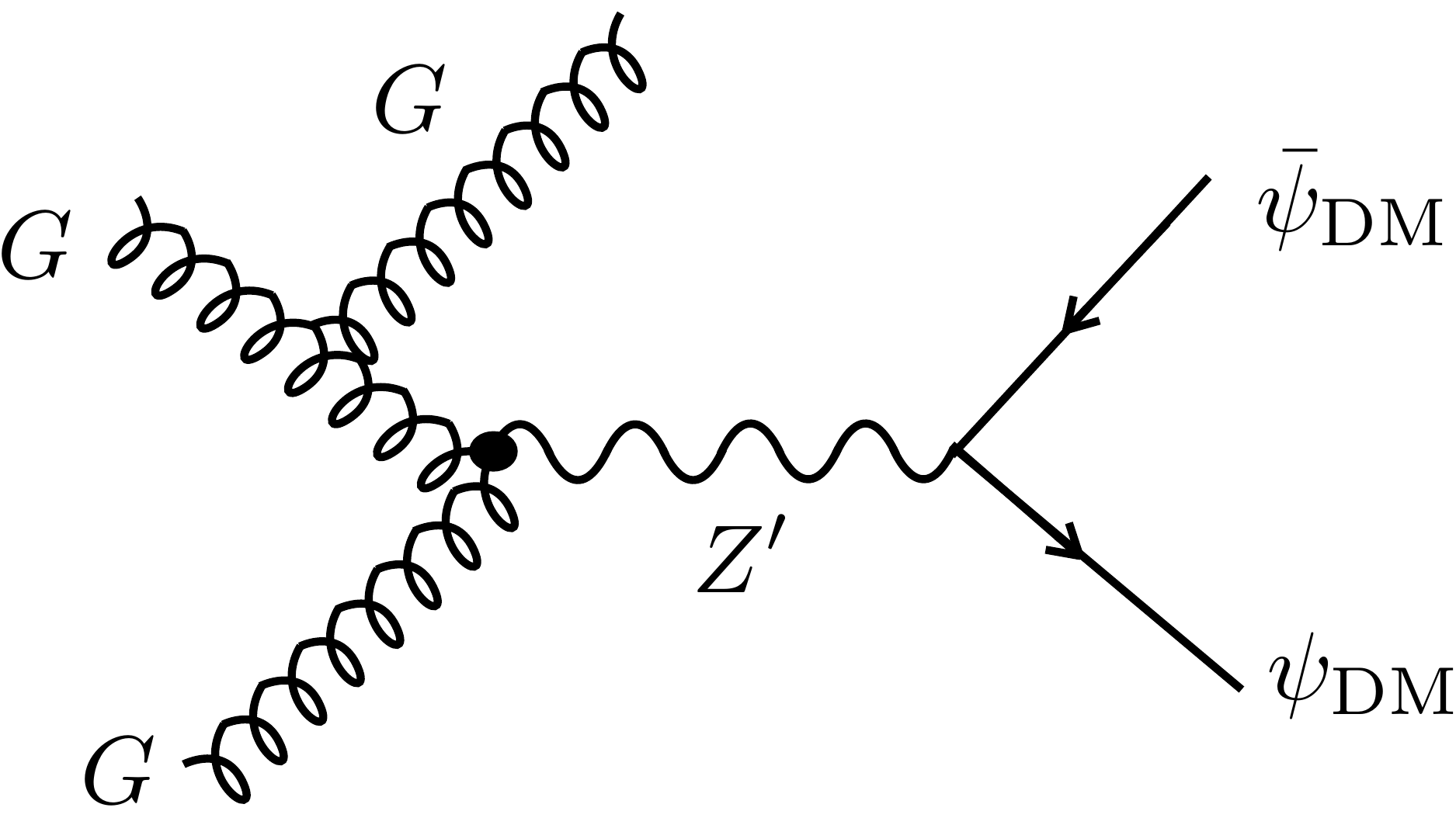}
\includegraphics[width=0.3\textwidth]{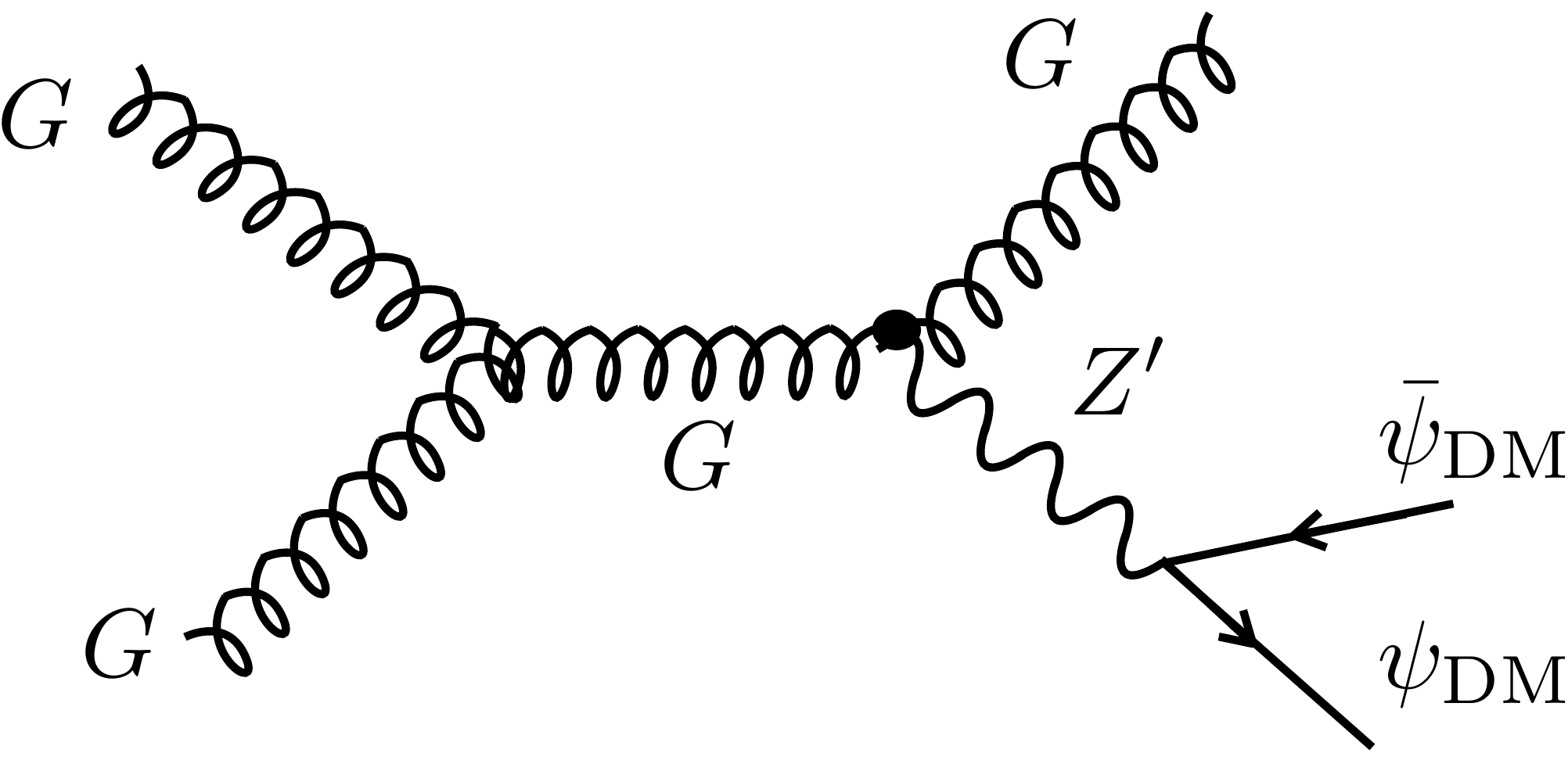}  
\includegraphics[width=0.3\textwidth]{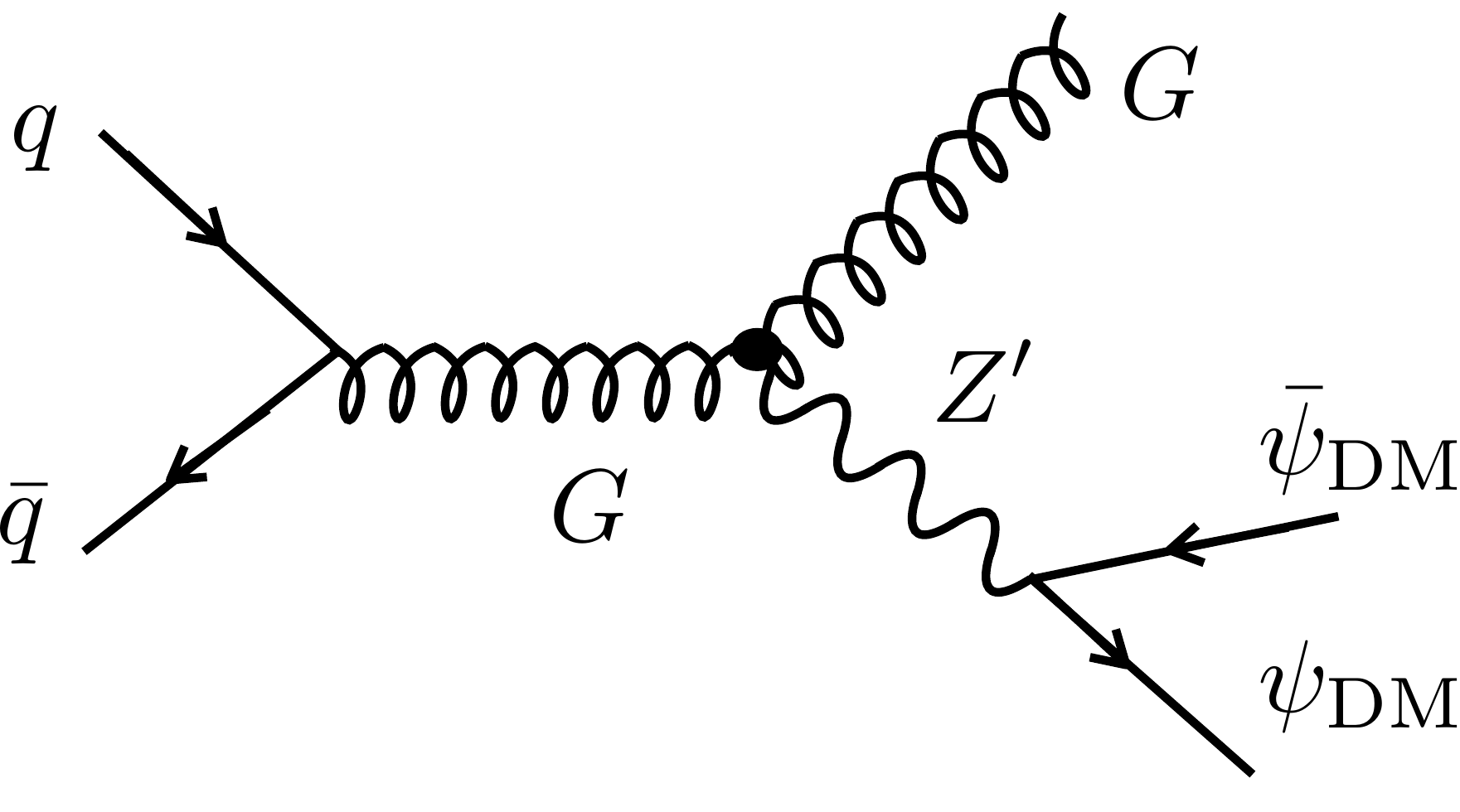}  
\end{center}
\caption{\label{fig:LHC} Dark matter production processes at the LHC (at partonic level), in association with 1 jet: $p~p\to j\bar\psi_{\rm DM}\psi_{\rm DM}$. }
\end{figure}

The model described in previous sections can be probed at the LHC. Indeed the $Z'$-gluon-gluon vertex makes possible to produce a dark matter pair out of two protons, provided a $Z'$ is produced. Typical production channels are shown in Fig. \ref{fig:LHC}, where we consider a generic process:
\begin{equation}
p ~p \to j ~\bar\psi_{\rm DM}~\psi_{\rm DM}
\end{equation}
of a proton-proton collision giving rise to 1 jet, plus missing energy ($E_T^{\rm miss}$). 

%%%%%%%%%%%%%%%%%%%%%%%%%%%%%%%%%%%%%%%%%%%%%%%%%%%%%%%%%%%%%%%%%%%%%%%%%%
\begin{figure}[htbp]
\begin{center} 
\includegraphics[width=0.7\textwidth]{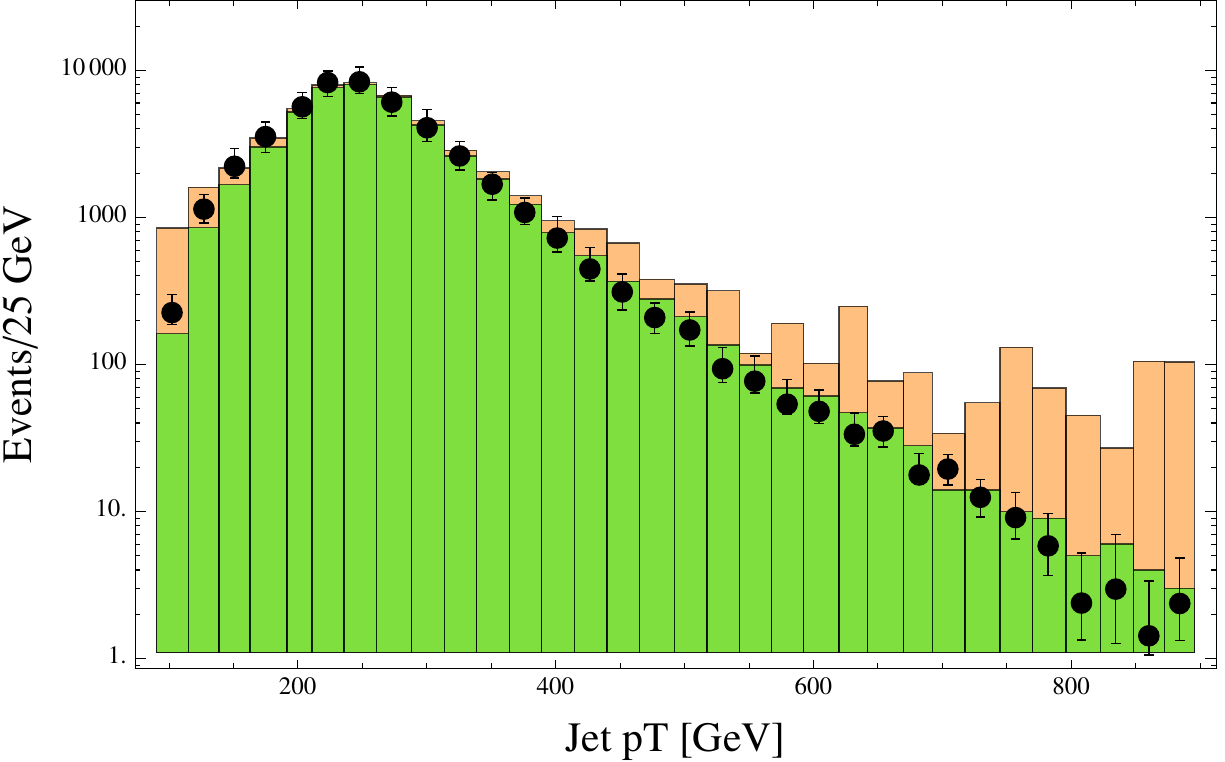}
\end{center}
\caption{\label{fig:histo} Histogram of $p_T^{\rm jet}$ corresponding to a particular choice of the model parameters (see text for details). The signal is shown in orange. 
The background (green bars) and data (points) are taken from the CMS analysis. }
\end{figure}

\begin{figure}[htbp]
\begin{center} 
\includegraphics[width=0.7\textwidth]{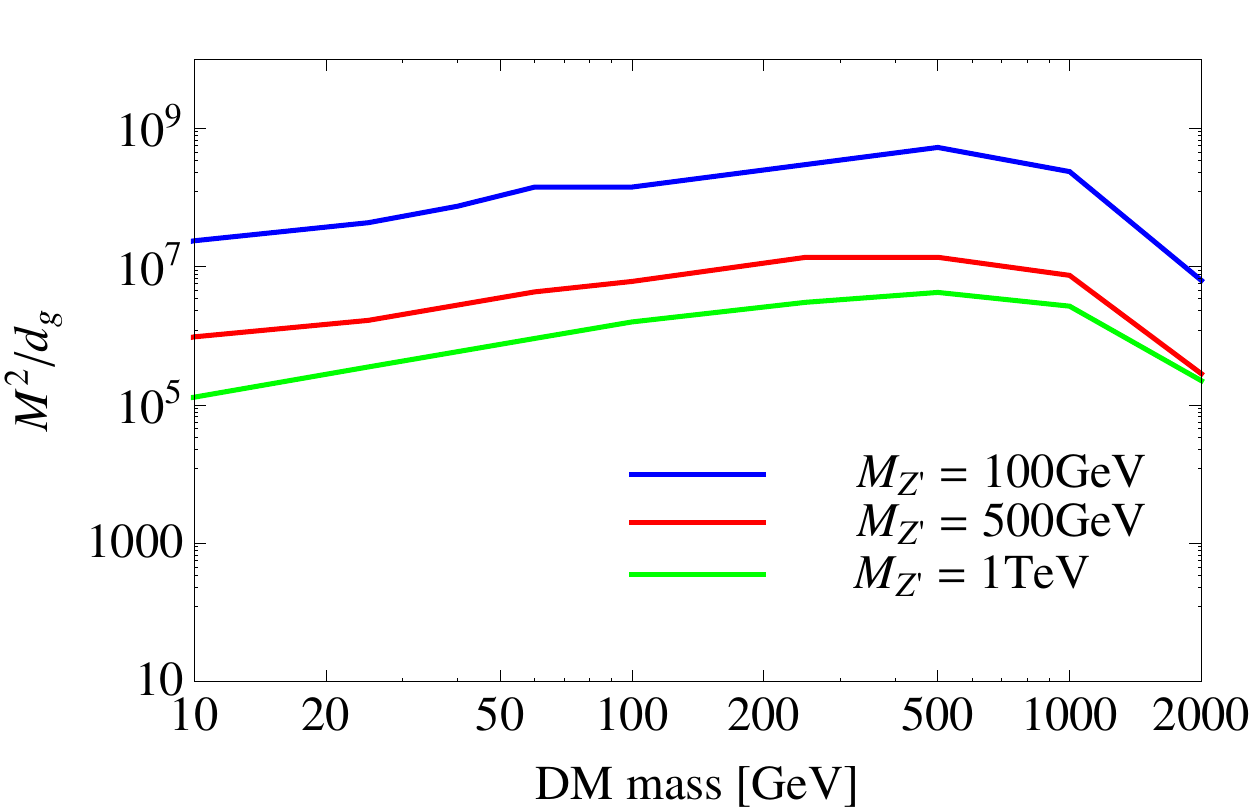}
\end{center}
\caption{\label{fig:LHC2} 90\% CL lower bounds on the quantity $M^2/d_g$ as a function of the dark matter mass, for $M_{Z'}=100$ GeV (blue), 500 GeV (red) 
and 1 TeV (green). Based on the CMS analysis with collected data using a center-of-mass energy of 8 TeV and a luminosity of 19.5/fb. }
\end{figure}

%%%%%%%%%%%%%%%%%%%%%%%%%%%%%%%%%%%%%%%%%%%%%%%%%%%%%%%%%%%%%%%%%%%%%%%%%%
\noindent
The monojet final state was first studied using Tevatron data \cite{Bai:2010hh} in the framework of effective $\psi_{\rm DM}$-quark interactions of different nature. In a
 similar fashion, bounds to dark matter effective models have been obtained by analyzing single-photon final states using LEP  \cite{Fox:2011fx} and LHC \cite{sanz} data. An interesting 
 complementarity between these two approaches has been analyzed in \cite{Mambrini:2011pw}. Since then, the ATLAS and CMS groups have taken the mono-signal 
 analyses as an important direction in the search for dark matter at the LHC (see \cite{ATLAS:2012zim} and \cite{CMS:rwa} for the most recent results from ATLAS and 
 CMS, respectively). The most important background to the dark matter signal is coming from the Standard Model production of a $Z$ boson decaying to a neutrino
  pair ($Z\to\bar\nu\nu$), however, in the inclusive analysis other processes like $W\to\ell\nu$ are considered as well. Other interesting and solid studies can be found
  in \cite{Goodman:2010yf}.
\newline\newline\noindent
In this paper we use the monojet data coming from the CMS analysis \cite{CMS:rwa}, which collected events using a center-of-mass energy of 8 TeV up to an integrated 
luminosity of 19.5/fb. We perform the analysis by looking at the distribution of the jet's transverse momentum ($p_T^{\rm jet}$), taking the background analysis given in
 \cite{CMS:rwa} and simulating on top the 
signal coming from our model. For the event generation we use {\tt CalcHEP.3.4.2} \cite{Belyaev:2012qa}. 
\newline\newline\noindent
A typical histogram is shown in Fig. \ref{fig:histo}, where we have used  $m_\psi=10$ GeV, $M_{Z'}=100$ GeV and\footnote{We took for the figure
 the illustrative case  $|X_L-X_R|g_X^2=1$. Results  other values of the coupling are obtained by a simple rescaling of the number of events.} $d_g/M^2 = 10^{-6}$ as the model parameters.
%%%%%%%%%%%%%%%%%%%%%%%%%%%%%%%%%%%%%%%%%%%%%%%%%%%%%%%%%%%%%%%%%%%%%%%%%%
%%%%%%%%%%%%%%%%%%%%%%%%%%%%%%%%%%%%%%%%%%%%%%%%%%%%%%%%%%%%%%%%%%%%%%%%%%
\newline\noindent
The results are shown in Fig. \ref{fig:LHC2}, where we show the exclusion power of the monojet analysis to the model. We present the bounds for
the quantity $M^2/d_g$ as a function of the dark matter mass, for three different values of the $Z'$ mass: 100 GeV, 500 GeV and 1 TeV. 
\newline\newline\noindent
The shape and relative size of the bounds can be understood by looking at the amplitude of the processes, which are proportional to $c^2 m_\psi^2/M_{Z'}^4$, where
 the coupling $c\equiv d_g/M^2$. For example,  given a $M_Z'$, for $m_\psi=10$ GeV the bounds are approximately 10 times weaker than those for 
 $m_\psi = 100$ GeV. However, for $m_\psi\gtrsim 1$ TeV the dark matter starts to be too heavy to be easily produced out of the 4 TeV protons, given the 
 PDF suppression of the quarks and gluons; so the DM production is close to be kinematically closed. On the other hand, for example at $m_\psi=100$ GeV,
  the bound for $M_{Z'}=100$ GeV is around 25 (100) times stronger than the one corresponding to $M_{Z'}=500 (1000)~$ GeV.

%%%%%%%%%%%%%%%%%%%%%%%%%%%%%%%%%%%%%%%%%%%%%%%%%%%%%%%%%%%%%%%%%%%%%%%%%%
%%%%%%%%%%%%%%%%%%%%%%%%%%%%%%%%%%%%%%%%%%%%%%%%%%%%%%%%%%%%%%%%%%%%%%%%%%

%===================================================================================================================
%==========================================   KINETIC MIXING    =========================================================
%===================================================================================================================

\subsection{Constraints on the kinetic mixing}

\begin{figure}[htbp]
\begin{center} 
\includegraphics[width=0.6\textwidth]{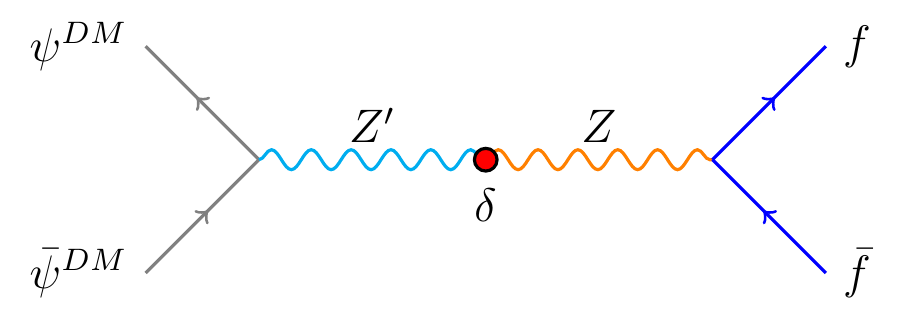}
\end{center}
\caption{\label{kinetic}\footnotesize{Example of $f\bar f$ production, from a dark matter annihilation and via an $s-$channel $Z/Z'$ exchange.}}
\end{figure}

\noindent
All through the analyses we considered a small kinetic mixing. However it can be interesting to check to what extent this hypothesis is valid.
Indeed, whereas it exists various constraints\footnote{The literature on the subject is very vast. We suggest for further reading \cite{Andreas:2012mt,Mambrini:2010dq} 
for dark matter
constraints, \cite{LHC} for LHC constraints, \cite{Javier} for string motivations and \cite{Gninenko:2013sr} for other studies. } on 
$\delta$ (from precision measurements, rare decay processes, $\rho$-parameter), a non-zero
kinetic mixing generates new annihilation diagrams ($s-$channel $Z/Z'$ exchange), as represented in Fig.\ref{kinetic}, which could modify our results\footnote{In all our study we use
the conventions described in \cite{Mambrini:2010dq}.}.

%%%%%%%%%%%%%%%%%%%%%%%%%%%%%%%%%%%%%%%%%%%%%%%%%%%%%%%%%%%%%%%%%%%%%%%%%%
%%%%%%%%%%%%%%%%%%%%%%%%%%%%%%%%%%%%%%%%%%%%%%%%%%%%%%%%%%%%%%%%%%%%%%%%%%

\noindent
To test the validity of our approach, we extract from Eq.(\ref{Eq:schannel}) an approximate solution for the gluonic annihilation cross section (we ignore here the
factors of $X_L-X_R$ for simplicity):

\begin{eqnarray}
\langle \sigma v \rangle_{GG} \simeq \frac{d_g^2}{M^4} \frac{2 g_X^4}{\pi} \frac{m^6_\psi}{M_{Z'}^4} \ . 
\end{eqnarray}

\noindent
Concerning the annihilation generated by the $s-$channel exchange of a $Z/Z'$ through kinetic mixings (see Fig. \ref{kinetic}), the expressions of the cross
section can be found in \cite{FIMP} and approximated by:\footnote{These expressions are valid in the regime $M_{Z'}>M_Z$ but a similar analysis can be performed
in the case $M_{Z'} < M_Z$.}

\begin{eqnarray}
&&
\langle \sigma v \rangle_{\delta} \simeq \frac{16}{\pi} g_X^2 g^2 \delta^2 \frac{m_\psi^2}{M_{Z'}^4} \qquad , \qquad m_\psi < M_Z
\nonumber
\\
&&
\langle \sigma v \rangle_\delta \simeq \frac{g_X^2 g^2 \delta^2 M_Z^4}{\pi m_\psi^2 M_{Z'}^4} \qquad  \quad , \qquad m_\psi > M_Z.
\end{eqnarray}

\noindent
We can then obtain the value of $\delta$ for which the process $\langle \sigma v \rangle_{\delta}$ dominates on $\langle \sigma v \rangle_{GG}$, invalidating our analysis done by ignoring the kinetic mixing :

\begin{eqnarray}
&&
\delta \gtrsim \frac{d_g}{M^2} \frac{g_X}{2\sqrt{2} g} m^2_\psi \qquad , \qquad 
m_\psi < M_Z
\nonumber
\\
&&
\delta \gtrsim \frac{d_g}{M^2} \frac{ \sqrt{2}g_X}{g} \frac{m^4_\psi}{M^2_Z}
\qquad , \qquad m_\psi > M_Z
\label{Eq:kinetic}
\end{eqnarray}

\noindent
which give for example for $m_\psi=200$ GeV and $g_X = 0.1$, $\frac{M^2}{d_g} \gtrsim \frac{10^4}{\delta}$ GeV$^2$. In other words, for values
of the coupling $\frac{d_g}{M^2} \lesssim 10^{-4} \times \delta \ GeV^{-2}$, the annihilation processes induced by kinetic mixing begin to compete with the
gluonic final state. Another interesting point is that the conditions are independent on the mass of the $Z'$ as soon as we assume $M_{Z'} \gg M_Z$.

\begin{figure}[htbp]
\begin{center} 
\includegraphics[width=0.4\textwidth]{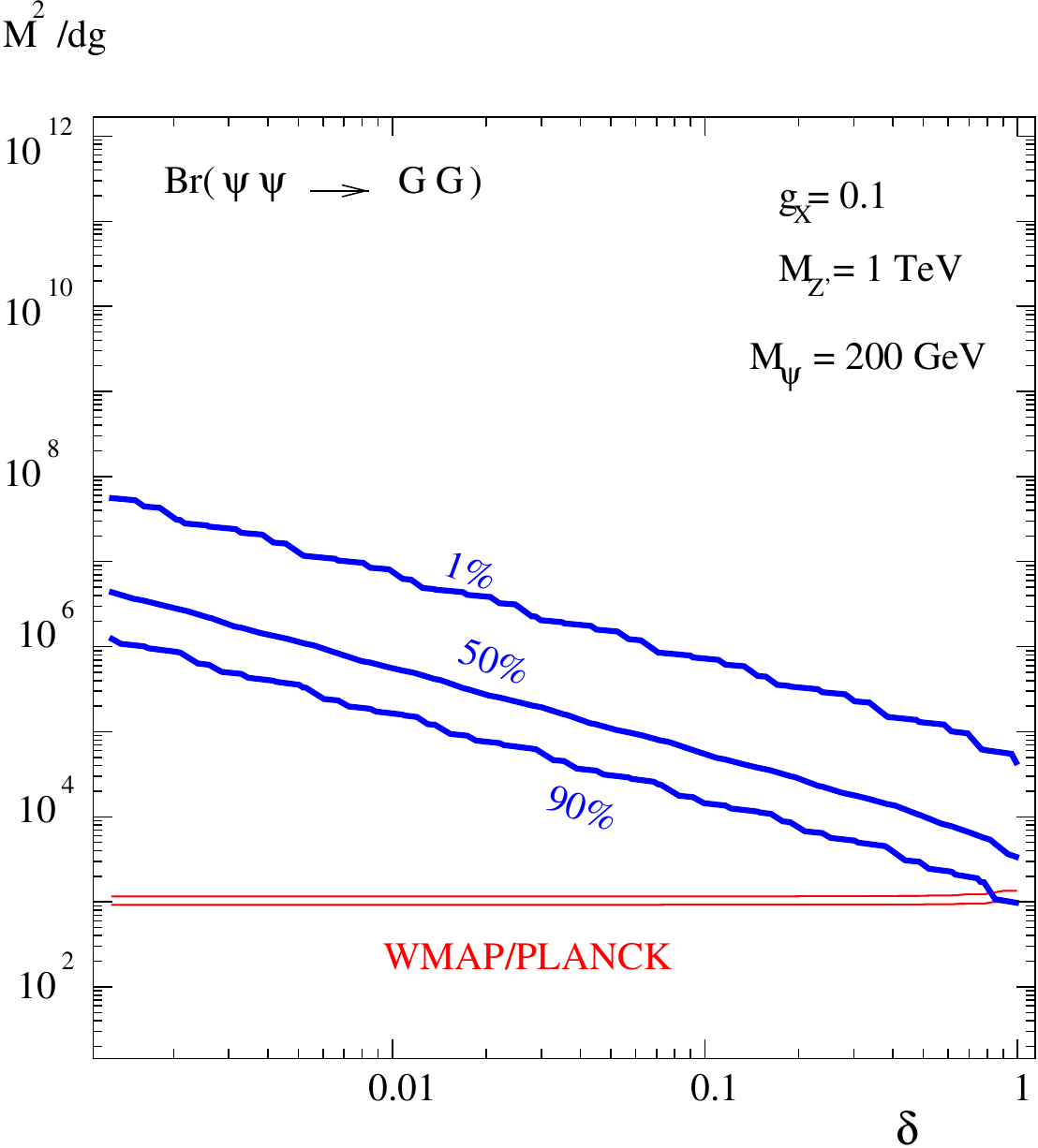}
\end{center}
\caption{\label{Fig:kinetic} \footnotesize{Gluonic branching fraction (blue line) of the annihilating dark matter in the plane ($\delta;d_g / M^2$) allowed by WMAP/PLANCK (red) data for a dark matter mass of 200 GeV, $g_X=0.1$ and $M_{Z'}$= 1 TeV.}}
\end{figure}

\noindent
To confirm our conclusions, we made a numerical analysis, allowing a non--zero kinetic mixing.
We show in Fig.(\ref{Fig:kinetic}) the iso-curve for the branching ratio $\langle \sigma v \rangle_{\psi \psi \rightarrow GG}$ in the plane ($\delta; d_g/M^2$) given
by our numerical analysis. We also draw the region allowed by WMAP at 
5$\sigma$~\footnote{The WMAP constraint is quite insensitive to $\delta$ in the range of values shown in Fig.(\ref{Fig:kinetic}),
 however for large $\delta$ and the same set of parameters we used, the dependence on $\delta$ becomes significant.}. 
 We took $M_{Z'}=1$ TeV, $m_{\psi}= 200$ GeV and $g_X=0.1$ but we checked
that the result is generic for broad regions in the parameter space\footnote{The helicity suppression of the dark matter 
annihilation into gluons plays an important role for this to happen.}. We first notice that the region respecting the cosmological bounds lie in a region where the gluonic fraction
is largely dominant (over 90\%). It is only for very high values of $\delta \simeq 0.8$ that the channel $\psi \psi \rightarrow Z/Z' \rightarrow$ SM SM can contribute
at a substantial level ($\simeq 10\%$) to the relic density computation, confirming with a surprising accuracy our analytic results Eq.(\ref{Eq:kinetic}).
 Such values for $\delta$ are already excluded by LEP experiments.

%=========================================================================================================
%======================================  SUMMARY CONSTRAINTS    ===========================================
%=========================================================================================================

%%%%%%%%%%%%%%%%%%%%%%%%%%%%%%
\subsection{Summary of the various constraints}

\begin{figure}[htbp]
\begin{center} 
\begin{tabular}{cc}
 \includegraphics[width=0.5\textwidth]{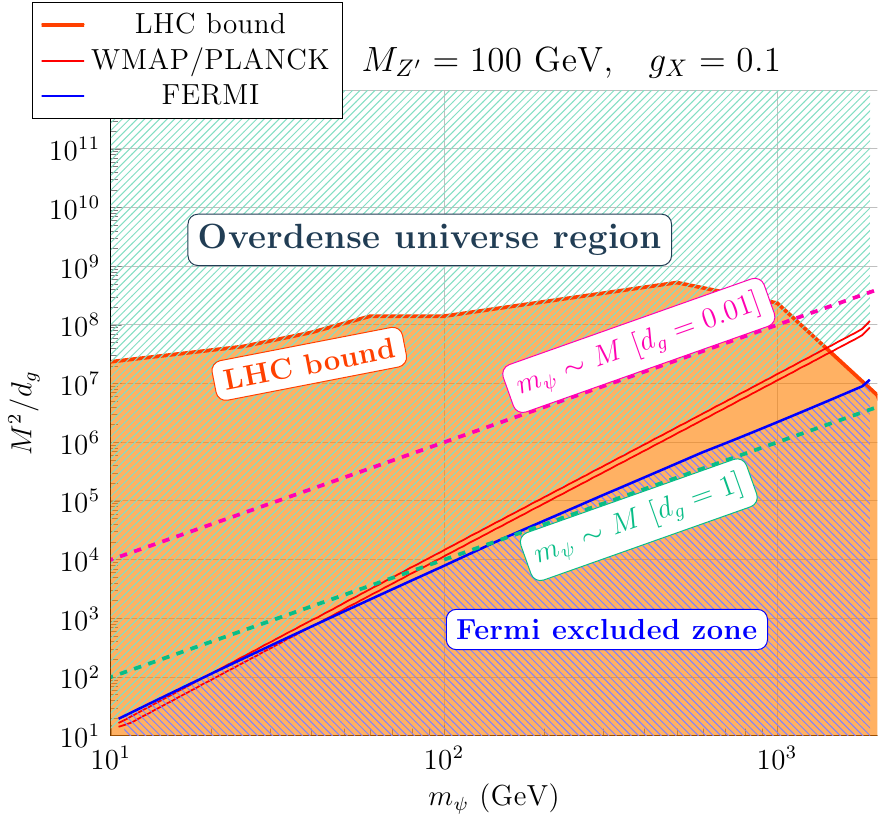}&\includegraphics[width=0.5\textwidth]{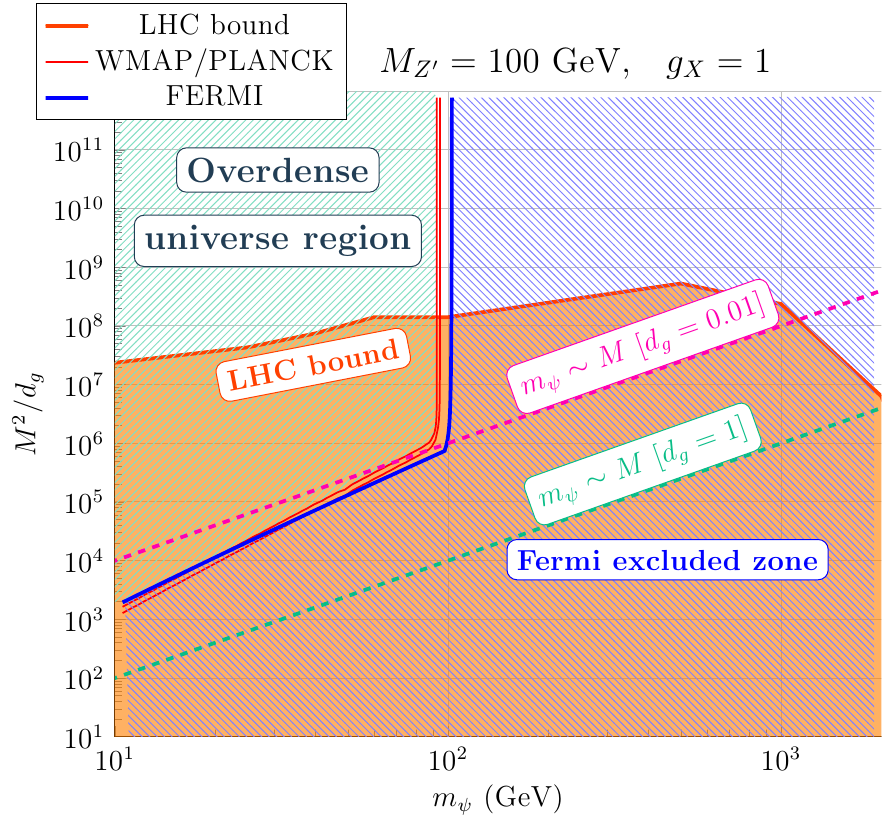}\\
 \includegraphics[width=0.5\linewidth]{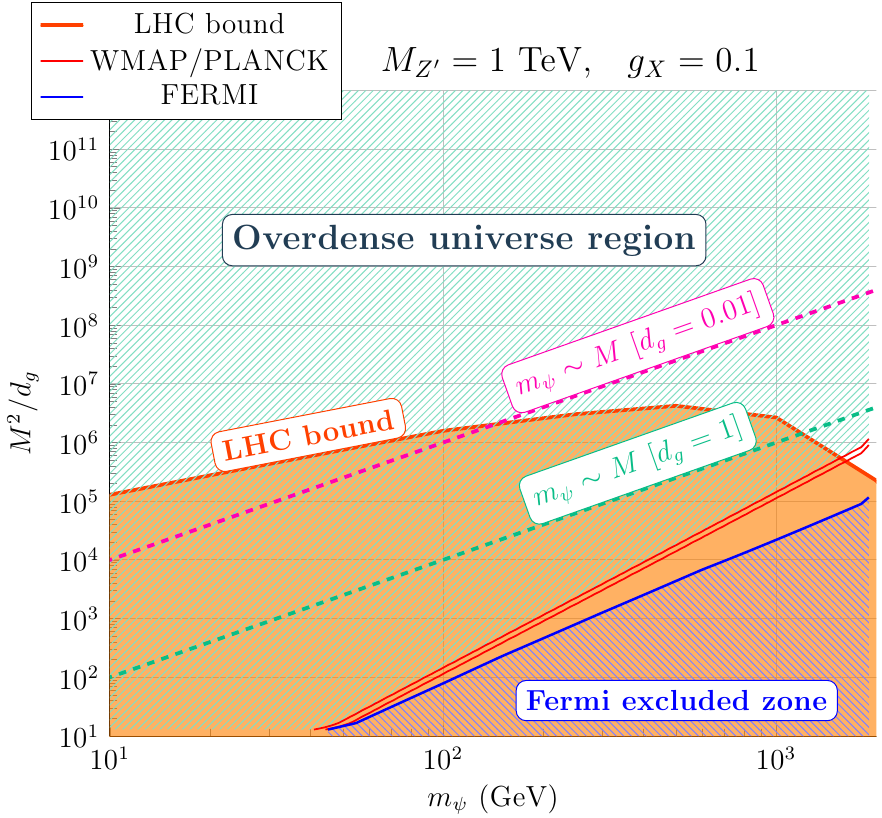}&\includegraphics[width=0.5\linewidth]{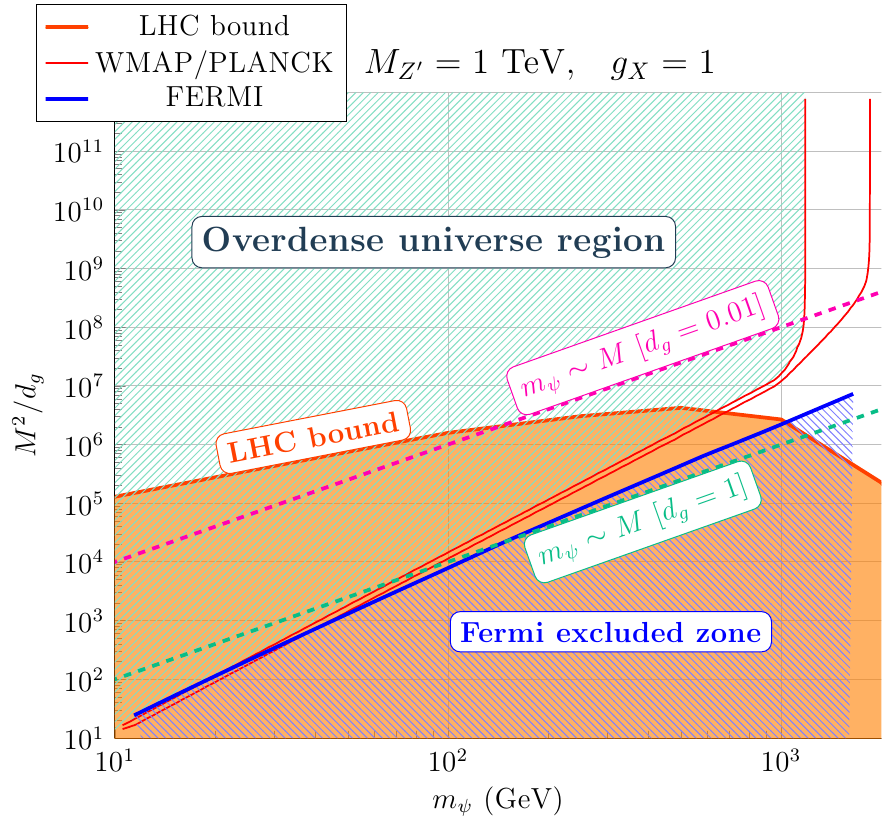}
\end{tabular}
\end{center}
\caption{\label{fig:synt} \footnotesize{Experimental constraints on $m_{\psi}$ and $M^2/d_g$ parameters, including LHC and universe overdensity constraints. Below the purple/dashed line $M \ll m_{\psi}$ and the effective theory analysis we made is not valid.}}
\end{figure}

\noindent
Now we can put together all the constraints we obtained on the parameter pair $(m_{\psi},\frac{M^2}{d_g})$ to see what are the new allowed regions in the
 parameter space. Superposing Fig.(\ref{Fig:WMAP}) and \ref{fig:LHC2}, we get a new representation of those {\em validity zones}, as represented in Fig.(\ref{fig:synt}).

%%%%%%%%%%%%%%%%%%%%%%%%%%%%%%%%%%%%%%%%%%%%%%%%%%%%%%%%%%%%%%%%%%%%%%%%%%
%%%%%%%%%%%%%%%%%%%%%%%%%%%%%%%%%%%%%%%%%%%%%%%%%%%%%%%%%%%%%%%%%%%%%%%%%%
 
\noindent
As explained earlier, parameters are allowed to lie below the red/full lines (Overdensity of the universe), above the orange/full line (LHC bounds on monojets production). 
Since the whole study has been released using effective dimension six operators generated by integrating out heavy fermions loops, one has to check that the
parameter range is still in the window where $M\gg m_{\psi}$. This is indicated on Fig.(\ref{fig:synt}) where we considered natural values of $d_g$ varying
between  $10^{-2}$ and $1$ (purple and green/dashed line,  respectively). Thus one can easily distinguish between the two regions
   $m_{\psi}\ll M$ (upper region) and $m_{\psi}\gg M$ (lower region).
   
 \noindent
In the case where $d_g\sim 10^{-2}$, it is important to notice that low values of the coupling constant $g_X$ provide almost no validity region in the parameter
 space since parameters have to lie above the purple/dashed line. On the other hand, for $g_X=1$ one can also notice that the allowed region is much larger
  in the case of a heavy $Z'$. The case $d_g\sim 1$ considerably relax the constraints since the validity zones are almost in the region where
   $m_{\psi}\ll M$ (below the green/dashed line).

%%%%%%%%%%%%%%%%%%%%%%%%%%%%%%%%%%%%%%%%%%%%%%%%%%%%%%%%%%%%%%%%%%%%%%%%%%%%%%%%%%%%%%%%%%%%%%
\section{$Z'$ annihilation into electroweak gauge bosons}

In the same way the $Z'$ boson couples to gluons via operators of dimension six, mediators with electroweak quantum numbers can generate operators 
coupling the $Z'$ boson to gauge bosons of the $SU(2) \times U(1)_Y$ electroweak sector. They can be parametrized as

\begin{eqnarray}
 \mc{L} &=& \frac{1}{M^2}\left\{ {D}^{\mu}\tta_X\left[ i(D^{\nu}H)^{\dagger}(c_1\tilde{F}^Y_{\mu\nu}+2c_2\tilde{F}^W_{\mu\nu})H+h.c.\right]\right.\ok
 &+&\left.\partial^mD_m \theta_X (d_1\mc{T}r(F^Y\tilde{F}^Y) +2d_2\mc{T}r(F^W\tilde{F}^W)) +
 d'_{ew} \partial^{\mu} D^{\nu} \theta_X {\rm Tr}(F_{\mu \rho}\tilde{F}^{\rho}_{\nu})  \nonumber
\right. \\
&+& \left. e_{ew}  D^{\mu} \theta_X {\rm Tr}(F_{\nu \rho} {\cal D}_{\mu} \tilde{F}^{\rho \nu})
+ e'_{ew} D_{\mu} \theta_X {\rm Tr}(F_{\alpha \nu} 
{\cal D}^{\nu} \tilde{F}^{\mu \alpha}) 
 \right\}\,. \label{ew1}
\end{eqnarray}
These effective operators give contributions to $Z'\rightarrow Z Z$, $Z'\rightarrow Z
\gamma$ and $Z'\rightarrow \g \g$ processes. We neglected such operators until now, since
they induce new free parameters in the model. They can contribute to SM matter production in the universe, which in turn can slightly relax our previous constraints on the parameter $\frac{d_g}{M^2}$.

Let us now consider the $Z'$ couplings to electroweak gauge bosons coming from the dimension-six operators $c_i$ and $d_i$ in (\ref{ew1}), by ignoring the others. The reason for ignoring the last ones $d',e$ and $e'$ is the same as for the gluonic
couplings. On the other hand, although beyond the goals of the present paper, we believe that the operators $c_i$ are induced and do contribute
in a computation with heavy loop of mediators, provided that part of mediator masses
come from couplings to the SM Higgs.   
The interaction lagrangian of the couplings $c_i,d_i$ to the electroweak sector are then given by

\begin{figure}[htbp]
\begin{center}
\includegraphics[width=14cm]{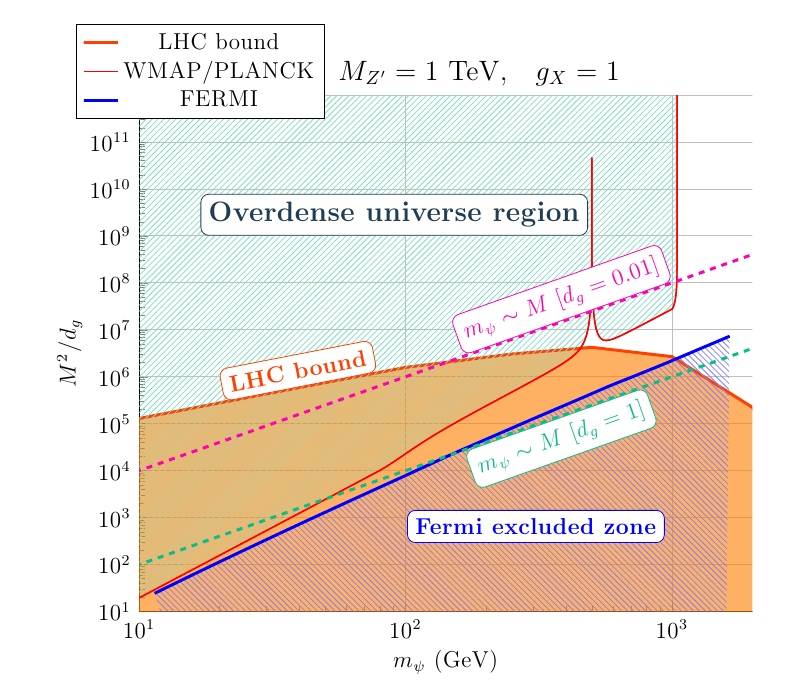}
\end{center}
\caption{\label{fig:figure} \footnotesize{Experimental constraints on the $(M^2/d_g, m_{\psi})$ parameters, taking into account dark matter couplings to all SM  gauge bosons and assuming $c_i=d_i=d_g$.}}
\end{figure}

\begin{itemize}
 \item [$\purple{\diamond}$] $Z' \rightarrow ZZ $ process :
 \begin{eqnarray}
 \Delta \mc{L}_{Z' \rightarrow ZZ} &=& g_Xm_{Z}v{\frac{\sin\tta_Wc_1+\cos \tta_Wc_2}{M^2}}\epsilon_{\m\n\rho\sigma}Z'^{\m} Z^{0 \nu}\partial^{\rho}Z^{0 \sigma}\ok
 &+&2{\frac{\sin^2\tta_Wd_1+\cos^2\tta_Wd_2}{M^2}}g_X\epsilon^{\m\n\rho\sigma} \partial^m Z'_{m}  \partial_{\m}{Z}_{\n}\partial_{\rho}{Z}_{\sigma}\, \ ,
\end{eqnarray} 
 \item [$\purple{\diamond}$] $Z' \rightarrow Z \gamma $ process :
 \begin{eqnarray}
 \Delta \mc{L}_{Z' \rightarrow Z \gamma} &=& g_Xm_{Z}v{\frac{\sin\tta_Wc_2-\cos\tta_Wc_1}{M^2}}\epsilon_{\m\n\rho\sigma}Z'^{\m} Z^{\nu 0}\partial^{\rho}A^{\sigma}\, \ok
 &+&4g_X\sin\tta_W\cos\tta_W{\frac{d_2-d_1}{M^2}}{\epsilon^{\m\n\rho\sigma}} \partial^m Z'_{m}\partial_{\m}{Z}_{\n} \partial_{\rho}A_{\sigma} \ , 
\end{eqnarray} 
\item [$\purple{\diamond}$] $Z' \rightarrow W^+W^-$ process :
 \begin{eqnarray}
 \Delta \mc{L}_{Z' \rightarrow W^+W^-} &=& g_Xv{\frac{c_2}{M^2}}Z'^{\m}{\epsilon_{\m\n\rho\sigma}} \ m_W (W^{\n -}\partial^{\rho}W^{+\sigma}+W^{\n +}\partial^{\rho}W^{-\sigma})\, \ok
 &+&4{\frac{d_2}{M^2}}g_X {\epsilon^{\m\n\rho\sigma}}\partial^m Z'_{m}\partial_{\m}{W^+}_{\n}\partial_{\rho}{W^-}_{\sigma} \ , 
\end{eqnarray} 
\item [$\purple{\diamond}$] $Z' \rightarrow \g\g$ process :
\begin{eqnarray}
 \Delta \mc{L}_{Z' \rightarrow \g\g} &=& 2{\frac{\cos^2\tta_Wd_1+\sin^2\tta_Wd_2}{M^2}}g_X\epsilon^{\m\n\rho\sigma} \partial^m Z'_{m}  \partial_{\m}{A}_{\n}\partial_{\rho}{A}_{\sigma}\, \ .
\end{eqnarray}
\end{itemize}
These interaction terms give rise to the cross sections for the s-channel displayed in
Appendix D.  They have to be added to the t-channel cross section. 
We can now add the resulting cross sections to the one of gluons production to consider a more precise constraint about universe overdensity, which is
\begin{equation}
 \langle(\sigma_{GG}+\sigma_{ZZ}+\sigma_{Z\gamma}+\sigma_{\gamma\gamma}+\sigma_{W^+W^-}) v\rangle_{s-channel}+\langle\sigma v\rangle_{t-channel} \geqslant \langle\sigma v\rangle_{thermal}\,.
\end{equation} 
Then,  assuming for simplicity that all the couplings appearing in the different six-dimensional operators are equal to $\frac{d_g}{M^2}$, which is a very strong hypothesis of course, we can plot a new constraint on this parameter, in a similar way we did before. This provides a new validity zone in the parameter space, as represented in Fig.\ref{fig:figure} (in the case where $M_{Z'}=1 \text{TeV}$ and $g_X=1$), in which we added the electroweak processes to the gluon couplings of Section 3. 

The resulting constraints are slightly relaxed, but the validity zones are not greatly enlarged, as anticipated earlier. One notice that the behaviour of the cross sections around $m_{\psi}=M_{Z'}/2$ is modified here, compared to the gluon production process. This happens because the electroweak gauge bosons $W^{\pm}$ and $Z^°$ are massive, unlike the gluons. Thus the Landau-Yang theorem does not apply and a real $Z'$ can be created, relaxing the constraints on $M^2/d_{g}$ parameter. Implications of Landau-Yang theorem can yet be extended to express some constraints on what kind of CP even operators can be written down to produce electro-weak gauge bosons; this has been done previously for $Z'\rightarrow Z,Z$ process in \cite{Keung:2008ve}. Our results are in agreement with theirs in the form of operators and resulting cross sections.

%%%%%%%%%%%%%%%%%%%%%%%%%%%%%%%%%%%%%%%%%%%%%%%%%%%%%%%%%%%%%%%%%%%%%%%%%%%%%%%%%%%%%%%%%%%%
%%%%%%%%%%%%%%%%%%%%%%%%%%%%%%%%%%%%%%%%%%%%%%%%%%%%%%%%%%%%%%%%%%%%%%%%%%%%%%%%%%%%%%%
{\bf Acknowledgements. }  We would like  to thank Massimo Bianchi, J.M. Moreno and V. Martin-Lozano for very useful discussions. E.D., L.H and Y.M 
thank the Galileo Galilei Institute for Theoretical Physics for the hospitality and the INFN for partial support during the 
completion of this work. This  work was supported in part by the European ERC Advanced Grant 226371 MassTeV, the 
French ANR TAPDMS ANR-09-JCJC-0146 the contract PITN-GA-2009-237920 UNILHC 
 and the Spanish MICINNs
Consolider-Ingenio 2010 Programme  under grant  Multi- Dark {\bf CSD2009-00064}.
Y.M.  acknowledges partial support from the European Union FP7 ITN INVISIBLES (Marie
Curie Actions, PITN- GA-2011- 289442), the ERC advanced grant Higgs@LHC. B.Z. acknowledges the support of MICINN, Spain, under the contract FPA2010-17747, as well as the hospitality of LPT, Orsay, during the completion of this project.

%%%%%%%%%%%%%%%%%%%%%%%%%%%%%%%%%%%%%%%%%%%%%%%%%%%%%%%%%%%%%%%%%%%%%%%%%%%%%%%%%%%%%%
\appendix 
\section{Gauge independence and unitary gauge}

In this Appendix we discuss the gauge independence of $Z'$ induced effective couplings. In the
Stueckelberg phase and after integrating out the heavy mediators, the effective 
action in  $R_{\xi}$ gauges is 
\begin{eqnarray}
&&{\cal L} = - \frac{1}{4} (F_{\mu \nu}^{Z'})^2 + \frac{1}{2} (\partial_{\mu} a_X - \frac{g_X}{2} V Z'_{\mu})^2 - \frac{1}{2 \xi} (\partial_{\mu} Z'^{\mu} + \xi 
\frac{g_X}{2} V a_X )^2 \nonumber \\
&& + Z'_{\mu} \Gamma^{\mu} (A) + a_X  \Gamma_a (A) - m_{\psi}
(e^{i a_X (X_L-X_R)/V} {\bar \psi}_L \psi_R + e^{- i a_X (X_L-X_R)/V} {\bar \psi}_R \psi_L  ) \ . \label{a1}
\end{eqnarray}
In (\ref{a1}), $\Gamma^{\mu} (A)$ describes the local (non-local) coupling between $Z'$ and SM gauge fields generated in the case where some heavy (light) fermions are charged under $Z'$. $\Gamma_a$ is the axionic coupling generated in this case by the heavy set of mediator fermions cancelling an
eventual gauge anomaly, which captures the low-energy remnant of the heavy mediator fermions in the infinite mass limit. 
Gauge invariance implies
\begin{equation}
\partial_{\mu} \ \Gamma^{\mu} (A) = \frac{g_X}{2} \ V \ \Gamma_a (A) \ .  \label{a2}
\end{equation}
At the abelian (three-point function) level, we can write
\begin{equation}
\Gamma_{\mu} \ = \ \frac{1}{2} \Gamma_{\mu \nu \rho} A^{\nu} A^{\rho} \quad , \quad
\Gamma^a \ = \ \frac{1}{2} \Gamma^a_{\nu \rho} A^{\nu} A^{\rho} \ , \label{a02}
\end{equation}
where $A^{\nu}$ denotes symbolically the SM gauge fields. 
As concrete examples, the operator $\Gamma_a $ coupling gluons to the axion is of the form $\Gamma_a \sim \Box \ {\cal T}r \ (G \tilde G) + 2 \ \partial_{\mu} \ {\rm Tr} (G_{\alpha \nu}
D^{\nu} {\tilde G}^{\mu \alpha})$ for the operators  induced by chiral but anomaly-free set of heavy mediators in Section 2.1, whereas is of the form 
$\Gamma_a \sim {\cal T}r \ (G \tilde G) $ for the anomalous sets of fermion mediators considered
in Section 2.2.  
In momentum space, the gauge invariance conditions for the three point function $Z' A A $ are 
\begin{eqnarray}
&& k_1^{\nu} \Gamma_{\mu \nu \rho} (k_i) \ = \ 0  \quad , \quad
k_2^{\rho} \Gamma_{\mu \nu \rho} (k_i) \ = \ 0  \nonumber \\
&& i (k_1+ k_2)^{\mu} \Gamma_{\mu \nu \rho} (k_i) \ = \ \frac{g_X}{2} V \Gamma^a_{\nu \rho} (k_i) \ . \label{a03} \\
\end{eqnarray} 
The $Z'$ and axion propagators are
\begin{equation}
\Delta_{\mu \nu}^{Z'} (q) = - i \frac{\eta_{\mu \nu} + (\xi-1) \frac{q_{\mu} q_{\nu}}{q^2 - \xi
M_{Z'}^2}}{q^2 - M_{Z'}^2} \quad , \quad \Delta_{a_X} (q) = \frac{i}{q^2 - \xi M_{Z'}^2} \  \label{a3} 
\end{equation}
and the unitary gauge corresponds to the limit $\xi \to \infty$. 
Whereas the issue of gauge-fixing  independence can be discussed in more general terms, we
prefer to analyse it in the relevant context for our work, fermions- 2 SM gauge fields interactions mediated by the $Z'$ exchange. In an arbitrary $R_{\xi}$ gauge, there are two contributions:
the $Z'$ and the axionic exchange:
\begin{eqnarray}
&& {\cal M} = {\bar v} (p_2)(-\frac{i g_X}{2}) [\frac{X_R +X_L}{2} \gamma^{\mu} + \frac{X_R-X_L}{2} \gamma^{\mu}
\gamma_5  ] u (p_1) \left(- i \frac{\eta_{\mu \nu} + (\xi-1) \frac{q_{\mu} q_{\nu}}{q^2 - \xi
M_{Z'}^2}}{q^2 - M_{Z'}^2} \right)  \Gamma^{\nu}  \nonumber \\ 
&& + {\bar v} \gamma_5 (X_L-X_R) \frac{m_{\psi}}{V} \frac{i}{q^2 - \xi M_{Z'}^2}  \Gamma_a 
u (p_1)\ , 
 \label{a4}
\end{eqnarray}
 where  $\Gamma^{\nu},\Gamma_a$ are the three-point functions coming from the operators present in (\ref{a1}), $q$ is the
 $Z'$ virtual momentum and $u (p), v (p)$ the Dirac spinors associated to the fermion (antifermion) $\Psi$ coupling to $Z'$, to be identified with the Dark Matter candidate in our paper. By using Dirac equation for the fermion $\Psi$ and the gauge invariance condition
 (\ref{a2}) in momentum space $ -i q _{\mu} \Gamma^{\mu} (k_i) = \frac{g_X}{2} V \Gamma_a (k_i)$, with
 $k_1,k_2$ the momenta of the two gauge bosons in the final space, we find   
\begin{eqnarray}
&& {\cal M} = {\bar v} (p_2)(-\frac{i g_X}{2}) \left[ \frac{X_R +X_L}{2} \gamma^{\mu} + \frac{X_R-X_L}{2} \gamma^{\mu}
\gamma_5  \right] u (p_1) \left(\frac{ -i \Gamma^{\mu}}{q^2 - M_{Z'}^2} \right) \nonumber \\ 
&& + \ {\bar v}(p_2) \gamma_5 (X_L-X_R) \frac{m_{\psi}}{V} \frac{i}{q^2 - M_{Z'}^2}  \Gamma_a u (p_1)
\ .  \label{a5}
\end{eqnarray}
As expected, due to gauge invariance, the $\xi$-dependence cancelled in the final result. Moreover,
the result can also be directly found in the unitary gauge with no axion field present. In this
case, the result is fully encoded in the unitary gauge computation
\begin{equation}
{\cal M} = {\bar v} (p_2)(- \frac{i g_X}{2}) \left[ \frac{X_R +X_L}{2} \gamma^{\mu} + \frac{X_R-X_L}{2} \gamma^{\mu}
\gamma_5  \right] u (p_1) \left(- i \frac{\eta_{\mu \nu} - \frac{q_{\mu} q_{\nu}}{M_{Z'}^2}}{q^2-M_{Z'}^2}
\right)  \Gamma^{\nu} \ . 
 \label{a6}
\end{equation}
 Notice that in the unitary gauge the lagrangian can be expressed entirely in terms
of 
\begin{equation}
{\tilde Z'}_{\mu} = Z'_{\mu} - \frac{2}{g_X V} \partial_{\mu} a_X \quad , \quad 
{\tilde \Psi}_{L,R} = e^{- \frac{i a_X }{V} X_{L,R}} \ \Psi_{L,R} \ . 
 \label{a7} 
\end{equation}
Unitary gauge captures correctly the result in the infinite mass limit of the heavy fermions.
For finite masses, there are corrections which are not captured by the naive unitary gauge 
computation.  
%%%%%%%%%%%%%%%%%%%%%%%%%%%%%%%%%%%%%%%%%%%%%%%%%%%%%%%%%%%%%%%%%%%%%%%%%%%%%%%%%%%%%%%%%%%%
\section{Three-point gauge boson amplitude and gauge effective action from heavy fermion loops: couplings to gluons}

In the case of CP invariance, the three-point gauge boson amplitude can be generally be written
as \cite{abdk}
\begin{eqnarray}
&& \Gamma^{\mu \nu \rho} = \epsilon^{\mu \nu \rho \alpha} ( A_1 k_{1 \alpha}
+ A_2 k_{2 \alpha}) + \nonumber \\
&& \left[  \epsilon^{\mu \nu \alpha \beta} (B_1 k_1^{\rho} + B_2 k_2^{\rho} )
+ \epsilon^{\mu \rho \alpha \beta} (B_3 k_1^{\nu} + B_4 k_2^{\nu} ) \right] 
\ k_{1 \alpha} k_{2 \beta} \ , \label{a8} 
\end{eqnarray}
where $A_i,B_i$ are Lorentz-invariant functions of the external momenta $k_i$. 
The functions $A_i$ which encode the generalized Chern-Simon terms (GCS) \cite{abdk} are superficially logarithmically divergent, whereas the functions $B_i$
are UV finite. However, $A_i$ are determined in terms of $B_i$ by using the Ward identities, 
which in case the heavy fermions form an anomaly-free set, are given by 
\begin{eqnarray}
 k_1^{\nu} \Gamma_{\mu \nu \rho} \ = \ 0 && \quad \rightarrow \quad A_2 = B_3 k_1^2 + B_4 k_1 k_2 \ , \nonumber \\
k_2^{\rho} \Gamma_{\mu \nu \rho} \ = \ 0 && \quad \rightarrow \quad
A_1 = B_2 k_2^2 + B_1 k_1 k_2 \ , \nonumber \\
 - (k_1+ k_2)^{\mu} \Gamma_{\mu \nu \rho} \ = \ && (A_1 - A_2) \
 \epsilon_{\nu \rho \alpha \beta} k_1^{\alpha}  k_2^{\beta} \ . \label{a08} 
\end{eqnarray}
The last current conservation is nontrivial in our case, since gauge invariance is realized
through an additional axionic coupling to gauge fields generated by heavy fermions, such that  we find (\ref{a03}). After comparison with (\ref{a08}), this implies 
\begin{equation}
\Gamma^a_{\nu \rho} \ = \ - \frac{2i}{g_X V} \ (A_1-A_2) \ \epsilon_{\nu \rho \alpha \beta} \ k_1^{\alpha} k_2^{\beta} \ . \label{a09} \\
\end{equation}
The situation here is different compared to the usual discussion of anomalies. The usual axionic couplings compensating triangle gauge anomalies are generated by chiral and non-anomaly free set of fermions. If the heavy fermions form an anomaly-free set, they do not generate such couplings, but dimension six operators for gauge fields and dimension seven axionic couplings, which cancel
between themselves their gauge variation. 
There are two contributions to $\Gamma^{\mu \nu \rho}$. The first is the triangle loop diagram
with no chirality flip/mass insertions, given by
\begin{equation}
\Gamma_{\mu \nu \rho}^{(1)} =  \sum_i t_{iaa} 
\int \frac{d^4 p}{(2 \pi)^4} \ {\rm Tr} \left[ \frac{\slp + {\slk}_2}{(p+k_2)^2 - M_i^2}
\gamma_{\rho}  \frac{\slp }{p^2 - M_i^2} \gamma_{\nu}
 \frac{\slp - {\slk}_1}{(p-k_1)^2 - M_i^2} \gamma_{\mu} \gamma_5 \right] \ . \label{a9}  
\end{equation}
where $t_{iaa} = {\rm Tr} (X_i T^a T^a)$.
There are also three other contributions with two mass insertions, of the type
\begin{equation}
\Gamma_{\mu \nu \rho}^{(2)} =  \sum_i t_{iaa}
\int \frac{d^4 p}{(2 \pi)^4} \ {\rm Tr} \left[ \frac{M_i}{(p+k_2)^2 - M_i^2}
\gamma_{\rho}  \frac{\slp }{p^2 - M_i^2} \gamma_{\nu}
 \frac{M_i}{(p-k_1)^2 - M_i^2} \gamma_{\mu} \gamma_5 \right] + \cdots \ ,  
 \label{a10} 
\end{equation}  
where $\cdots$ denote two similar contributions with the mass insertions permuted among
the three propagators. By using a Feynman parametrization and after performing a shift
of the momentum integral $p \to p + \beta k_1 - \alpha k_2$, we find
\begin{equation}
\Gamma_{\mu \nu \rho}^{(1)} = 2  \sum_i t_{iaa,L-R}
\int_0^1 d \alpha \int_0^1 d \beta \int \frac{d^4 p}{(2 \pi)^4} \ 
\frac{N_{\mu \nu \rho} (p,k_i)}{[ p^2 + \alpha (1-\alpha) k_2^2 +
 \beta (1-\beta) k_1^2 + 2 \alpha \beta k_1 k_2 - M_i^2]^3} \ ,  \label{a11} 
\end{equation}
where $t_{iaa, L-R} = {\rm Tr} [(X_L-X_R) T_a T_a]_i$ and where
\begin{eqnarray}
&& N_{\mu \nu \rho} (p,k_i) = {\rm Tr} \ \{ [\slp  + \beta {\slk}_1 +
(1-\alpha){\slk}_2 ] \gamma_{\rho} [\slp  + \beta {\slk}_1 
-\alpha {\slk}_2 ] \gamma_{\nu} [\slp - (1- \beta) {\slk}_1 
-\alpha {\slk}_2 ] \gamma_{\mu} \gamma_5 \} = \nonumber \\
&& - {\rm Tr} \ \{ \slp  \gamma_{\rho} \slp  [ (1- \beta) {\slk}_1 
+ \alpha {\slk}_2 ] \gamma_{\mu} \gamma_5 \} +
{\rm Tr} \ \{ [\beta {\slk}_1 +
(1-\alpha){\slk}_2 ] \gamma_{\rho} \slp   \gamma_{\nu} \slp  \gamma_{\mu} \gamma_5 \}
\label{a12}  \\
&& + {\rm Tr} \ \{ \slp  \gamma_{\rho} [\beta {\slk}_1 
-\alpha {\slk}_2 ] \gamma_{\nu} \slp  \gamma_{\mu} \gamma_5 \} - 
{\rm Tr} \ \{ [\beta {\slk}_1 +
(1-\alpha){\slk}_2 ] \gamma_{\rho} [\beta {\slk}_1 
-\alpha {\slk}_2 ] \gamma_{\nu} [(1- \beta) {\slk}_1 
+ \alpha {\slk}_2 ] \gamma_{\mu} \gamma_5 \} \nonumber 
\end{eqnarray}
The first three terms in (\ref{a12}) contribute to the ambiguous $A_i$ functions which will be however uniquely determined by the Ward identities (\ref{a08}). The last one, on the other hand,
is contributing to $B_i$ and equals
\begin{eqnarray}
&& {\rm Tr} \ \{ [\beta {\slk}_1 +
(1-\alpha){\slk}_2 ] \gamma_{\rho} [\beta {\slk}_1 
-\alpha {\slk}_2 ] \gamma_{\nu} [(1- \beta) {\slk}_1 
+ \alpha {\slk}_2 ] \gamma_{\mu} \gamma_5 \} = \nonumber \\ 
&& - 4 i  \ \{ [ \beta (2 \alpha +
\beta -1) k_{1 \rho} + \alpha (2- 2 \alpha - \beta) k_{2 \rho} ] \epsilon_{\mu \nu \alpha
\beta} k_1^{\alpha} k_2^{\beta} \nonumber \\
&& - \beta [(1-\beta)k_{1 \mu} + \alpha k_{2 \mu}] \epsilon_{\nu \rho \alpha \beta} k_1^{\alpha} k_2^{\beta}  + \beta [(1-\beta)k_{1 \mu} +
\alpha k_{2 \mu}] \epsilon_{\rho \mu \alpha \beta} k_1^{\alpha} k_2^{\beta}  \nonumber \\
&&  - \epsilon_{\mu \nu \rho \alpha}  [\beta^2 k_1^2 - \alpha (1-\alpha)
k_2^2 + (1- 2 \alpha) \beta k_1 k_2 ] [(1-\beta) k_1^{\alpha} + \alpha  k_2^{\alpha} ] \}
\label{a13}
\end{eqnarray} 
Integrating over the internal momentum $p$ and over the Feynman parameters $\alpha, \beta$
one finally finds
\begin{eqnarray}
&& \Gamma_{\mu \nu \rho}^{(1)} = - \sum_i  \frac{i t_{iaa, L-R}}{48 \pi^2 M_i^2} 
\{ (4 k_1 + k_2)_{\rho} \epsilon_{\mu \nu \alpha \beta} - 
(2 k_1 + 3 k_2)_{\mu} \epsilon_{\nu \rho \alpha \beta} +
(2 k_1 + 3 k_2)_{\nu} \epsilon_{\rho \mu \alpha \beta} \} k_1^{\alpha} k_2^{\beta} 
\nonumber \\
&& + A-terms \ , \label{a14} 
\end{eqnarray}
where the A terms in (\ref{a8}) are determined at the end by the Ward identity (\ref{a08}). The last step is the symmetrization in the two gluonic legs, which leads to the final result
\begin{eqnarray}
&& \Gamma_{\mu \nu \rho}^{(1) symm.} = - \sum_i  \frac{i t_{iaa, L-R}}{48 \pi^2 M_i^2} 
\{ (7 k_1 + 3 k_2)_{\rho} \epsilon_{\mu \nu \alpha \beta} - 
5 ( k_1 +  k_2)_{\mu} \epsilon_{\nu \rho \alpha \beta} +
(3 k_1 + 7 k_2)_{\nu} \epsilon_{\rho \mu \alpha \beta}  \} k_1^{\alpha} k_2^{\beta} 
\nonumber \\
&& + \cdots =  - \sum_i  \frac{i t_{iaa, L-R}}{12 \pi^2 M_i^2} 
\{ (- k_{1 \rho} \epsilon_{\mu \nu \alpha \beta} + 
2 ( k_1 +  k_2)_{\mu} \epsilon_{\nu \rho \alpha \beta} -
k_{2 \nu} \epsilon_{\rho \mu \alpha \beta} \} k_1^{\alpha} k_2^{\beta}  + 
{\rm A-terms} \  \ ,  \label{a15}
\end{eqnarray}
where in order to find the last line we used the identities
\begin{eqnarray}
&& (\epsilon^{\nu \rho \alpha \beta} k_1^{\mu} +
\epsilon^{\rho \mu \alpha \beta} k_1^{\nu} +
\epsilon^{\mu \nu \alpha \beta} k_1^{\rho} ) \ k_{1\alpha}  k_{2\beta}
=  \epsilon^{\mu \nu \rho \alpha} (k_1^2 k_{2 \alpha} - k_1 k_2 k_{1 \alpha}) \ , \nonumber \\
&& (\epsilon^{\nu \rho \alpha \beta} k_2^{\mu} +
\epsilon^{\rho \mu \alpha \beta} k_2^{\nu} +
\epsilon^{\mu \nu \alpha \beta} k_2^{\rho} ) \ k_{1\alpha}  k_{2\beta}
=  \epsilon^{\mu \nu \rho \alpha} (k_1 k_2 k_{2 \alpha} - k_2^2 k_{1 \alpha}) \ . 
\label{a16}
\end{eqnarray}
The contribution with two mass insertions $\Gamma_{\mu \nu \rho}^{(2)}$ are easily seen to give
terms correcting the coefficients $A_i$ in (\ref{a8}). As such, they are fixed 
by the Ward identities (\ref{a08}).  
The complete three-point function, including the $A_i$ coefficients defined in (\ref{a8}), is then given by
\begin{equation}
\Gamma_{\mu \nu \rho}^{\cal O} =  - \sum_i  \frac{i t_{iaa, L-R}}{12 \pi^2 M_i^2} 
\{ [ 2 ( k_1 +  k_2)_{\mu} \epsilon_{\nu \rho \alpha \beta} - k_{1 \rho} \epsilon_{\mu \nu \alpha \beta} - k_{2 \nu} \epsilon_{\rho \mu \alpha \beta} ] k_1^{\alpha} k_2^{\beta} 
+ \epsilon_{\mu \nu \rho \alpha} k_1 k_2 (k_2-k_1)^{\alpha} \} \ .   
\label{a18}
\end{equation}
Notice that (\ref{a18}) can be cast in the general form (\ref{a8}). Indeed, by using identities of the type (\ref{a16}), one can also write
\begin{equation}
\Gamma_{\mu \nu \rho}^{\cal O} =   \sum_i  \frac{i t_{iaa, L-R}}{12 \pi^2 M_i^2} 
\{ [ (3 k_{1 \rho} + 2 k_{2 \rho}) \epsilon_{\mu \nu \alpha \beta} + 
(2 k_{1 \nu} + 3 k_{2 \nu}) \epsilon_{\rho \mu \alpha \beta} ] k_1^{\alpha} k_2^{\beta} + \epsilon_{\mu \nu \rho \alpha} 
[ (2 k_1^2+ 3k_1 k_2) k_2^{\alpha} - (2 k_2^2+ 3k_1 k_2) k_1^{\alpha}] \} \ ,   
\label{a018}
\end{equation}
from which the coeff. $A_i,B_i$ in (\ref{a8}) can be readily identified. 
The final result for the $Z'$ couplings  is then described by the operator
\begin{equation}
{\cal O} \ = \ \frac{g_3^2 }{24 \pi^2 } \sum_i \ {\rm Tr} \left(\frac{(X_L-X_R) T_a T_a}{M^2}\right)_i
\ \left[ \partial^{\mu} D_{\mu} \theta_X {\cal T}r (G {\tilde G}) - 2
D_{\mu} \theta_X {\rm Tr}(G_{\alpha \nu} {\cal D}^{\nu} \tilde{G}^{\mu \alpha})  \right] \ . 
\label{a17}
\end{equation}
The antisymmetric part of (\ref{a14}), which is relevant if one replaces $Z'$ by another gluon, can be shown to vanish, by using the identities  (\ref{a16}). Therefore, one-loops of heavy mediators do not generate triple SM gauge boson vectors operators of the type (\ref{g7}) and there are no new phenomenological constraints coming from purely SM contact operators.  

%%%%%%%%%%%%%%%%%%%%%%%%%%%%%%%%%%%%%%%%%%%%%%%%%%%%%%%%%%%%%%%%%

\section{Vanishing of the operator $\mathcal{T}r(F^X F_{SM} \tilde{F}_{SM})$ and a useful identity.}

Here we show that the operator $\mathcal{T}r(F^X F_{SM} \tilde{F}_{SM})$ is identically zero. The proof is the same for any SM gauge field, so we consider
the gluons for definiteness. 
In the unitary gauge, the $Z'$-gluon-gluon vertex coming from this operator is
proportional to
\begin{equation}
\frac{1}{M^2} \epsilon^{\lambda\nu\rho\sigma}(\partial_{\mu}Z'_{\nu}\partial^{\mu}G^A_{\lambda}\partial_{\rho}G^A_{\sigma} - \partial_{\mu}Z'_{\nu}\partial_{\lambda}G^A_{\mu}\partial_{\rho}G^A_{\sigma}-\partial_{\nu}Z'_{\mu}\partial^{\mu}G^A_{\lambda}\partial_{\rho}G^A_{\sigma} + \partial_{\nu}Z'_{\mu}\partial_{\lambda}G^A_{\mu}\partial_{\rho}G^A_{\sigma}) 
\end{equation}
In momentum space, denoting by $k_1,k_2$ the momenta of the two gluons, the linearized (abelian) $Z' GG$ vertex, after symmetrization of the two gluons, is given by 
\begin{equation}
\Gamma^{\mu \nu \rho} = 
\epsilon^{\nu \rho \sigma \tau} k_{1 \tau} k_2^{\mu} k_2^{\sigma} +
\epsilon^{\nu \rho \mu \sigma } (k_1k_2 k_{1\sigma} - k_1^2 k_{2\sigma}) +
\epsilon^{\rho \mu \sigma \tau} k_1{\nu} k_{1 \sigma} k_2^{\tau}
- \epsilon^{\nu \mu \sigma \tau} k_1{\rho} k_{1 \sigma} k_2^{\tau} \ . 
\end{equation}
Its vanishing can be seen by starting from the identity
\begin{equation} 
( \epsilon^{\nu \rho \sigma \tau} k_3^{\mu} + \epsilon^{\rho \mu \sigma \tau} k_3^{\nu} + \epsilon^{\mu \nu \sigma \tau} k_3^{\rho} ) \ k_2^{\sigma}
k_1^{\tau} \ = \ \epsilon^{\mu \nu \rho \tau} \ ( k_2k_3 k_{1 \tau}
 - k_1 k_3 k_{2 \tau} )  \ . \label{identity}
\end{equation}
The identity is actually valid for any vector $k_3$, that can be chosen, as in (\ref{a16}), to be one of the gluon momenta $k_{1,2}$, or the $Z'$ momentum $k_3 = - (k_1+k_2)$. 

If the linearized abelian  part of the operator vanishes, it has
to completely vanish because of gauge invariance.

%%%%%%%%%%%%%%%%%%%%%%%%%%%%%%%%%%%%%%%%%%%%%%%%%%%%%%%%%%%%%%%%%%%%%%%%
\section{The s and t-channel dark matter annihilation cross sections}

\subsection{The s-channel electroweak annihilation cross sections into electroweak gauge bosons}
The interaction terms of coeff. $c_i,d_i$ in (\ref{ew1}) give rise to the following cross sections for the s-channel
\begin{itemize}
 \item [$\purple{\diamond}$] $Z' \rightarrow ZZ $ process :
 \begin{eqnarray}
 &&\sigma_{\psi^{DM},\psi^{DM} \rightarrow Z,Z}={\left(  \frac{\sin\tta_Wc_1+\cos \tta_Wc_2}{M^2}\right)^2}\frac{v^2 g_X^4\left(s-4 m_{Z}^2\right)}{\left(M_{Z'}^2-s\right)^2+M_{Z'}^2\Gamma(Z')^2}\sqrt{\frac{s-4 m_{Z}^2}{s-4 m_{\psi}^2}} \times \ok
 && \frac{M_{Z'}^4(s-4 m_{Z}^2)(X_L^2+X_R^2)(2 m_{\psi}^2 + s)+ m_{\psi}^2 (X_L-X_R)^2 (6m_{Z}^2(s-M_{Z'}^2)^2-3 M_{Z'}^4 (s-4m_{Z}^2))}{768 \pi   M_{Z'}^4 s} \ , \ok
 \end{eqnarray}
 \item [$\purple{\diamond}$] $Z' \rightarrow \gamma Z $ process :
 \begin{eqnarray}
 &&\sigma_{\psi^{DM}\psi^{DM} \rightarrow \gamma Z}=\frac{\theta (s-m_{Z}^2)g_X^4}{\Gamma(Z') ^2 M_{Z'}^2+\left(M_{Z'}^2-s\right)^2}\sqrt{\frac{s}{s-4m_{\psi}^2}}\ok
  &\times&\left({ \sin^2\tta_W\cos^2\tta_W\frac{(d_2-d_1)^2}{M^4}} \frac{  m_{\psi}^2 ({X_L}-{X_R})^2(s-m_{Z}^2)^3(s-M_{Z'}^2)^2}{4\pi M_{Z'}^4}+{v^2\frac{(\sin\tta_W c_2 - \cos\tta_Wc_1)^2}{M^4}}\times\right.\ok
 &&\frac{ (m_{Z}^2-s)^3 \left(M_{Z'}^4 (2m_{\psi}^2+s) (m_{Z}^2+s)(X_L^2+X_R^2)+m_{\psi}^2 (X_L-X_R)^2 \left(-6 m_{Z}^2 M_{Z'}^2 s+3 m_{Z}^2 s^2-3 M_{Z'}^4 s\right)\right)}{768 \pi M_{Z'}^4 s^3 }\ok
 &-&{m_{Z}v\frac{(\sin\tta_W c_2 - \cos\tta_Wc_1)(d_2-d_1)}{M^4}}\times\left.\frac{  m_{\psi}^2 ({X_L}-{X_R})^2(s-m_{Z}^2)^3(s-M_{Z'}^2)^2}{8\pi M_{Z'}^4}\right) \ , 
  \end{eqnarray}
 \item [$\purple{\diamond}$] $Z' \rightarrow \gamma \gamma $ process :
 \begin{eqnarray}
\sigma_{\psi^{DM}\psi^{DM} \rightarrow \gamma\gamma}={\frac{(\cos^2 \tta_W d_1 +\sin^2 \tta_W d_2)^2}{M^4}}\frac{(-s+M_{Z'}^2)^2}{(-s+M_{Z'}^2)^2+ M_{Z'}^2\Gamma(Z')^2}\frac{ g_X^4 m_{\psi}^2 s^2 ({X_L}-{X_R})^2}{32 \pi  M_{Z'}^4} \sqrt{\frac{s}{s-4m_{\psi}^2}} \ . 
\ok
  \end{eqnarray}
  Notice the vanishing of the cross-section for the on-shell $Z'$ case $s=M_{Z'}^2$,
  in agreement with the Landau-Yang theorem \cite{landauyang}.
 \item [$\purple{\diamond}$] $Z' \rightarrow W^+W^- $ process :
 \begin{eqnarray}
 &&\sigma_{\psi^{DM}\psi^{DM} \rightarrow W^+W^-}= \frac{\theta (s-4 m_W^2)\left(s-4 m_W^2\right)^{3/2}g_X^4}{ \left(\Gamma ^2 M_{Z'}^2+\left(M_{Z'}^2-s\right)^2\right)\sqrt{s-4 m_{\psi}^2}}\ok
 &\times&\left(  {\left(\frac{d_2}{M^2}\right)^2}\frac{ m_{\psi}^2 s  \left(M_{Z'}^2-s\right)^2(X_L-X_R)^2}{16 \pi  M_{Z'}^4 }+{\left(\frac{c_2}{M^2}\right)}{\left(\frac{d_2}{M^2}\right)}\frac{vm_W\left(M_{Z'}^2-s\right)^2  \left(  m_{\psi}^2  (X_L-X_R)^2\right)}{16 \pi  M_{Z'}^4 } \right.\ok
 &+& {\left(\frac{c_2 v}{M^2}\right)^2}\left.\left(\frac{ (s-4 m_W^2)(X_L^2+X_R^2)(2 m_{\psi}^2 + s)}{384 \pi  s  }+\frac{ m_{\psi}^2 (X_L-X_R)^2 (6m_W^2(s-M_{Z'}^2)^2-3 M_{Z'}^4 (s-4m_W^2))}{384 \pi M_{Z'}^4 s  }\right)\right)\,.\ok
\end{eqnarray}
\end{itemize}

%%%%%%%%%%%%%%%%%%%%%%%%%%%%%%%%%%%%%%%%%%%%%%%%%%%%%%%%%%%%%%%%%%
\subsection{The t-channel dark matter annihilation into $Z'Z'$}

We give here the exact formula of the t-channel process cross-section as a function of the center of mass energy squared $s$ :

\begin{eqnarray}
 &&\langle\sigma v\rangle_{t-ch.}=\frac{g_X^4 v}{1024 \pi ^2 M_{Z'}^4 s }\sqrt{\frac{s-4 M_{Z'}^2}{s-4 m_{\psi}^2}} \left\{-2 m_{\psi}^2 (4 M_{Z'}^2-s) (X_L-X_R)^4-8 M_{Z'}^4 \left(X_L^4+X_R^4\right)\right.\ok
 &+&\frac{8 \coth^{-1}\left(\frac{2 M_{Z'}^2-s}{  \sqrt{(s-4 m_{\psi}^2) (s-4 M_{Z'}^2)}}\right)}{\left(2 M_{Z'}^2-s\right) \sqrt{(s-4 m_{\psi}^2)(s-4 M_{Z'}^2)} } \times\ok
 &&\left[m_{\psi}^4 \left(2 M_{Z'}^4 (3 X_L-X_R) (X_L+X_R)^2 (X_L-3 X_R)+4 M_{Z'}^2 s (X_L-X_R)^4-s^2 (X_L-X_R)^4\right)\right.\ok
 &+&2 m_{\psi}^2 M_{Z'}^2 \left(4 M_{Z'}^4 \left(-2 X_L^4+X_L^3 X_R-2 X_L^2 X_R^2+X_L X_R^3-2 X_R^4\right)+s^2 (X_L-X_R)^2 \left(X_L^2+X_R^2\right)\right.\ok
 &+&\left. \left. 2 M_{Z'}^2 s \left(-3 X_L^4+4 X_L^3 X_R+2 X_L^2 X_R^2+4 X_L X_R^3-3 X_R^4\right)\right)+ 2 M_{Z'}^4 \left(4 M_{Z'}^4+s^2\right) \left(X_L^4+X_R^4\right) \right]\ok
 &-&\frac{4 M_{Z'}^4}{m_{\psi}^2 \left(s-4 M_{Z'}^2\right)+M_{Z'}^4} \left[m_{\psi}^4 \left(X_L^2-6 X_L X_R+X_R^2\right)^2\right.\ok
 &+&\left.\left. 2 m_{\psi}^2 \left(M_{Z'}^2 \left(-3 X_L^4+6 X_L^3 X_R+2 X_L^2 X_R^2+6 X_L X_R^3-3 X_R^4\right)+s \left(X_L^2-X_R^2\right)^2\right)+2 M_{Z'}^4 \left(X_L^4+X_R^4\right)\right]\right\}\ok
\end{eqnarray}

%%%%%%%%%%%%%%%%%%%%%%%%%%%%%%%%%%%%%%%%%%%%%%%%%%%%%%%%%%%%%%%%%%%%%%%%%%%%%%%

\pagebreak
\nocite{}
%\bibliography{bmn}
\bibliographystyle{unsrt}

\end{document}